\documentclass[aip,cha,reprint]{revtex4-1}

\usepackage{color}
\usepackage{graphicx} 
\usepackage{xfrac} 
\usepackage{amssymb, amsmath, amsthm, amsfonts}

\newcommand{\BE}{\begin{equation}}
\newcommand{\EE}{\end{equation}}
\newcommand{\BA}{\begin{eqnarray}}
\newcommand{\EA}{\end{eqnarray}}
\newcommand{\bx}{{\bf x}}
\newcommand{\bv}{{\bf v}}
\newcommand{\mP}{\mathbf{P}}

\begin{document}


\title{Flow networks: A characterization of geophysical fluid transport} 



\author{Enrico Ser-Giacomi}
\author{Vincent Rossi}
\author{Crist\'{o}bal L\'{o}pez}
\author{Emilio Hern\'{a}ndez-Garc\'{\i}a}
\affiliation{$^1$IFISC (CSIC-UIB), Instituto de F\'{\i}sica
Interdisciplinar y Sistemas Complejos, Campus Universitat de
les Illes Balears, E-07122 Palma de Mallorca, Spain.}


\date{February 2, 2015}

\begin{abstract}
We represent transport between different regions of a fluid
domain by {\sl flow networks}, constructed from the discrete
representation of the Perron-Frobenius or transfer operator
associated to the fluid advection dynamics. The procedure is
useful to analyze fluid dynamics in geophysical contexts, as
illustrated by the construction of a flow network associated to
the surface circulation in the Mediterranean sea. We use
network-theory tools to analyze the flow network and gain
insights into transport processes. In particular we
quantitatively relate dispersion and mixing characteristics,
classically quantified by Lyapunov exponents, to the degree of
the network nodes. A family of {\sl network entropies} is
defined from the network adjacency matrix, and related to the
statistics of stretching in the fluid, in particular to the
Lyapunov exponent field. Finally we use a network community
detection algorithm, \textit{Infomap}, to partition the
Mediterranean network into coherent regions, i.e. areas
internally well mixed, but with little fluid interchange
between them.
\end{abstract}


\maketitle 


\begin{quotation}
Water and air transport are among the basic processes shaping
the climate of our planet. Heat and salinity fluxes change sea
water density, and thus drive the global thermohaline
circulation. Atmospheric winds force the ocean motion, and also
transport moisture, heat or chemicals, impacting the regional
climate. These considerations of geophysical fluid motion
suggests viewing fluid transport as a transportation network in
which fluid advances along different branches that eventually
split and merge. In this paper we exploit this interpretation
of fluid transport as a flow network so that we can use the
powerful techniques of modern network theory to better
characterize transport, mixing and dispersion, with examples
from ocean flows.
\end{quotation}


%
%

%

\section{Introduction}
\label{sec:intro}

The last two decades have seen important advances in the
Lagrangian description of transport and mixing in fluid flows
driven by concepts from dynamical systems theory. Nowadays the
techniques used can be roughly classified as follows. On the
one hand, some approaches focus on the geometric objects
--lines, surfaces-- separating fluid regions with different
properties. These geometric objects are often identified with
invariant manifolds
\cite{RomKedar1990,Poje1999,Mancho2006,Mancho2008,Balasuriya2012},
and more generally they are known under the name of {\sl
Lagrangian Coherent Structures}\cite{Haller2000,Peacock2010}.
Recent advances identify them as minimally stretching material
lines\cite{Haller2012}. On the other hand, another class of
algorithms have focussed on computing stretching-like fields in
the fluid domain, such as different types of Lyapunov exponents
or other Lagrangian
descriptors\cite{Haller2001,Joseph2002,dOvidio2004,Shadden2005,Huhn2012,Mancho2013}.
Ridges or singular lines in such fields turn out to be related,
under suitable conditions, to the Lagrangian Coherent
Structures of the previous approach, and thus organize the
fluid flow. Finally, there is a line of research focussing on
the moving fluid regions themselves, the so-called set-oriented
methods
\cite{Froyland2003,Froyland2005,Froyland2007,Dellnitz2009,
Froyland2010,Santitissadeekorn2010,Levnajic2010,Froyland2012Three,Tallapragada2013}.
The relationships among the different approaches have been
discussed in the
literature\cite{Lekien2007,Froyland2009,Branicki2010,Haller2011,
*Farazmand2012Erratum,Karrasch2012,Froyland2012Finite}. The
geometric approaches are designed to follow specific structures
during particular transport events, whereas the coarse-graining
inherent to the set-oriented methods makes them useful also to
estimate statistical properties in more extended space and time
intervals. Stretching-field methods can be used to follow
particular events or, by simple
averaging\cite{Waugh2008,Rossi2009,Hernandez-Carrasco2012},
also to characterize dispersion and stirring statistics in
large areas or long times.

One of the basic tools in the set-oriented methods is the
Perron-Frobenius or transfer operator, which quantifies the
amount of fluid transported from some initial region to other
ones under time evolution. A discretized version of that
operator is a transport matrix indicating which part of the
fluid domain is connected with which one, and by what amount of
flow. In this matrix form, the transfer operator can be read as
an adjacency matrix that defines a transportation or flow
network, an analogy that has been recently
recognized\cite{Preis2004,Dellnitz2006,Santitissadeekorn2007,
Nilsson-Jacobi2012,Speetjens2013,Thomas2014,Rossi2014}. Within
this network interpretation, the powerful tools of network or
graph
theory\cite{Albert2002,Dorogovtsev2002,Boccaletti2006,Newman2010}
become available to extract information about the transport
processes.

Network approaches have been used for geophysical systems in
the context of climate networks
\cite{Tsonis2006,Yamasaki2008,Donges2009b} in which the
connections among the different locations represent statistical
relationships between climatic time series from these
locations, inferred from correlations and other statistical
methods
\cite{Malik2011,Gozolchiani2011,Barreiro2011,Berezin2012}.
There is some recent work relating the connectivity given by
correlations to underlying fluid flow
\cite{Molkenthin2014networks}. In this Paper we analyze
directly the network describing the material fluid flow among
different locations, which we call \emph{flow network} or
\emph{transport network}. Among other characteristics this
network is directed, weighted\cite{Newman2004Analysis},
spatially embedded\cite{Barthelemy2011} and
time-dependent\cite{Holme2012}. We illustrate the general ideas
with an exemplary network derived from a realistic simulation
of the surface flow in the Mediterranean sea. Our focus is here
on the description with network tools of two important aspects
of transport, namely the quantification of \emph{dispersion}
and \emph{mixing}, and the identification of \emph{coherent
regions} which remain relatively isolated from neighboring
fluid. Relationships are drawn with the previous approaches
that used the geometric, stretching field, and set-oriented
methodologies described above, in particular with Lyapunov
exponents and with almost-invariant sets. In most of the paper
we use the language of ``water flow" appropriate to our ocean
dynamics example, but our methodology is in fact equally
applicable to atmospheric motions, to other fluid-dynamics
settings and even to flows in the phase space of more abstract
dynamical
systems\cite{Froyland2003,Dellnitz2006,Santitissadeekorn2007}.

The paper is organized as follows. After this introduction we
describe the basic steps to construct a flow network from fluid
velocity data (Sect. \ref{sec:flownet}). We apply them to a
surface flow field modelled for the Mediterranean sea. in Sect.
\ref{sec:Mediterranean}. The resulting network is studied in
Sect. \ref{subsec:network:dispersion} to characterize
dispersion and mixing in different regions. Appendix A
complements some technical aspects relevant here. In Sect.
\ref{subsec:network:coherent} we apply the network community
detection method \textit{Infomap} (described in more detail in
Appendix B) to identify coherent regions in the sea, well mixed
internally but with little exchange among them. The paper
finishes with a Conclusions section.

\section{Flow network construction from fluid motion}
\label{sec:flownet}

Since fluid flow is a process occurring in continuous space, a
discretization procedure involving a coarse-graining of space
is needed to have access to the techniques of network theory.
Advantages of the discrete point of view have already been
shown in geophysical contexts
\cite{Froyland2007,Dellnitz2009,Santitissadeekorn2010,Froyland2012Three,Froyland2014}.
Here we enumerate the steps needed to construct the discrete
transport network starting from the continuous flow.

\subsection{Discretization of the fluid domain: nodes}
\label{subsec:flownet:discretization}

Networks are composed of discrete building blocks:
\emph{nodes}. Being fluid flow a continuous system we need a
discretized version of it to give a network representation. To
do this we subdivide the fluid domain of interest in a large
number $N$ of boxes, $\{B_i, i=1,...,N\}$, so that network node
$j$ represents the fluid box $B_j$. Although it is not strictly
necessary, we consider here the case in which boxes have the
same area (in twodimensional flows) or volume (for three
dimensions). Then each box will contain exactly the same amount
of fluid.

\subsection{Lagrangian simulation: links and weights}
\label{subsec:flownet:Lagrangian}

To complete the construction of our transport network, we need
to establish the connections between nodes (i.e. boxes in the
fluid domain). We establish a directional \emph{link} between
two nodes when an exchange of fluid occurred from one to the
another during a given time interval. The \emph{weight} of this
link will be proportional to the amount of fluid transported.
This quantity could be obtained from a Lagrangian point of view
by following trajectories of ideal fluid particles and keeping
record of their initial and final positions (i.e. starting and
ending nodes) during the time interval considered.

More formally we integrate for a fixed time $\tau$ the equation
of motion for each particle, from initial condition $\bx_0$ at
time $t_0$ until the final position $\bx$ at $t_0+\tau$, using
a velocity field $\bv(\bx,t)$. This defines the {\sl flow map}
$\Phi_{t_0}^\tau$:
\BE
\bx(t_0+\tau)=\Phi_{t_0}^\tau (\bx_0)
\EE
which moves around single fluid particles. By considering the
action of the flow map on all the points contained in a fluid
region $A$ we define the action of $\Phi_{t_0}^\tau$ on whole
sets: $A(t_0+\tau)=\Phi_{t_0}^\tau (A(t_0))$.

\subsection{Construction of the network adjacency matrix}
\label{subsec:flownet:adjacency}

Applying the flow map to the discrete boxes, we will have an
estimation of the flow among each pair of nodes. More
explicitly, given the collection of boxes $\{B_i, i=1,...,N\}$,
we represent the transport between them by the discrete version
of the Perron-Frobenious operator $\mP(t_0,\tau)$, obtained
within the Ulam approach, whose matrix elements are given
by\cite{Froyland2003,Froyland2005,Froyland2007,Dellnitz2009,Froyland2010,Santitissadeekorn2010}:
\BE
\mP(t_0,\tau)_{ij} = \frac{m\left(B_i \cap
\Phi_{t_0+\tau}^{-\tau}(B_j)\right)}{m(B_i)} \ .
\label{PF}
\EE
$m(A)$ is a measure assigned to the set $A$. In our case it is
the amount of fluid it contains, i.e. simply its area or
volume. Other measures referring for example to heat or salt
content could be implemented for future applications. Eq.
(\ref{PF}) states that the flow from box $B_i$ to box $B_j$ is
the fraction of the contents of $B_i$ which is mapped into
$B_j$. We refer to the figure in Appendix A for a plot of the
different sets involved. If a nonuniform distribution of some
conserved tracer is initially released in the system such that
$\{p_i(t_0), i=1,...,N\}$ is the amount of such tracer in each
box $\{B_i\}$ at the initial instant $t_0$, the matrix
$\mP(t_0,\tau)$ gives the evolution of this distribution after
a time $\tau$ as $p_j(t_0+\tau)=\sum_{i=1}^{N}
p_i(t_0)\mP(t_0,\tau)_{ij}$. Writing the $\{p_i\}$ as row
vectors: $p(t_0+\tau)=p(t_0)\mP(t_0,\tau)$. A probabilistic
interpretation of Eq. (\ref{PF}) is that $\mP(t_0,\tau)_{ij}$
is the probability for a particle to reach the box $B_j$, under
the condition that it started from a uniformly random position
within box $B_i$. The matrix $\mP(t_0,\tau)$ is row-stochastic,
i.e. it has non-negative elements and $\sum_{j=1}^{N}
\mP(t_0,\tau)_{ij}=1$, but not exactly column stochastic. The
quantity $\sum_{i=1}^{N} \mP(t_0,\tau)_{ij}$ measures the ratio
of fluid present in box $B_j$ after a time $\tau$ with respect
to its initial content at time $t_0$. This ratio will be unity,
and the matrix doubly stochastic, if the flow $\bv(\bx,t)$ is
incompressible.

\begin{figure}
\centering
\includegraphics[width=\columnwidth, clip=true]{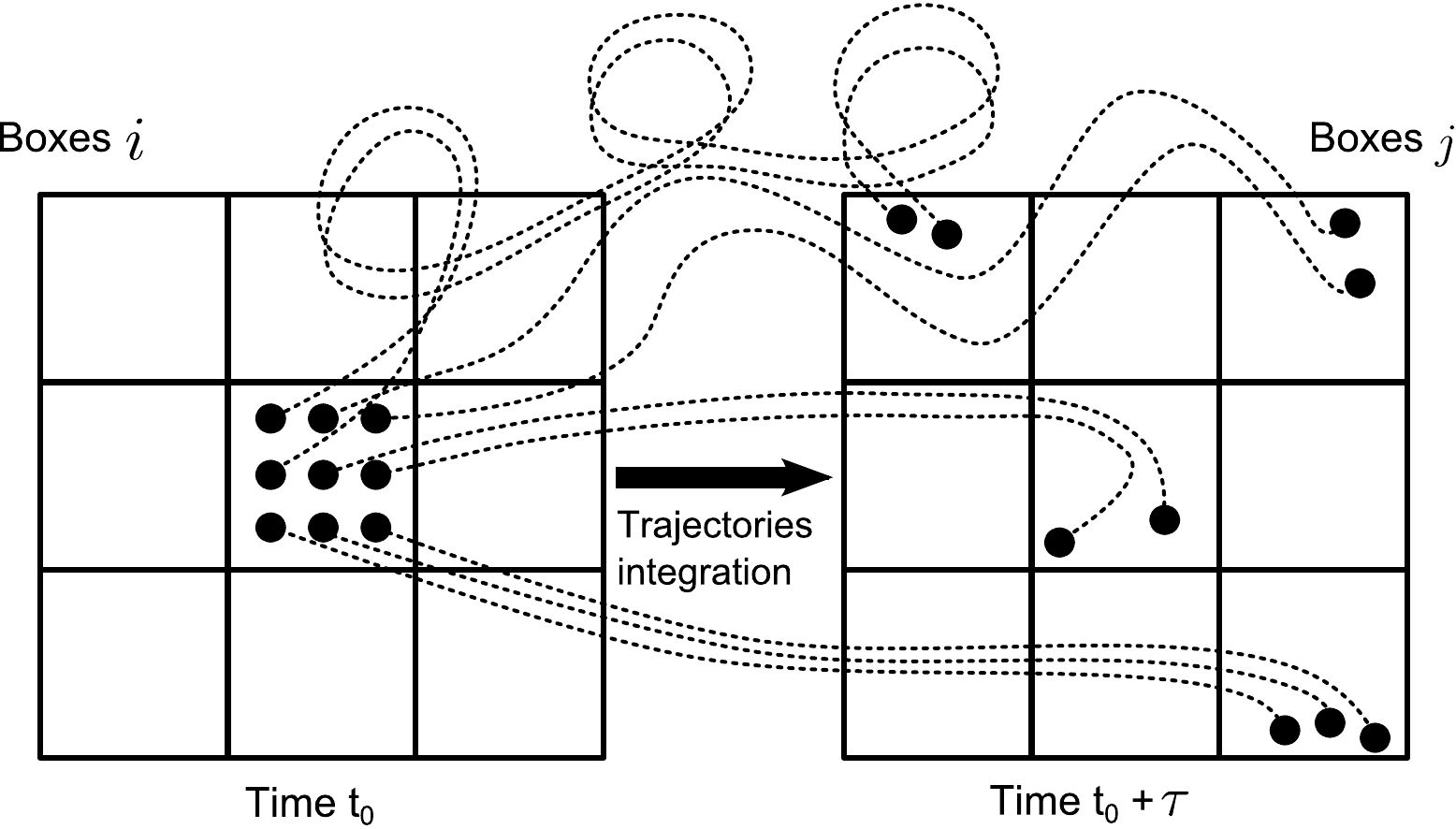}
\caption{Transport matrix construction from tracer's advection, following
Eq. (\ref{PFparticles}).
}
\label{fig:transmatrix}
\end{figure}

As a standard way to evaluate numerically the matrix in Eq.
(\ref{PF}) we apply the Lagrangian map to a large number of
particles released uniformly inside each of the boxes $\{B_i,
i=1,...,N\}$ (see Fig. \ref{fig:transmatrix}). The initial
number of particles $N_i$ in each box, a proxy of the amount of
fluid it contains, should be proportional to its measure
$m(B_i)$ which, with our choice of equal area or volume,
results in seeding the same number of particles in each box.
The number of particles transported from box $B_i$ to box $B_j$
gives an estimation of the flow among these boxes, and a
numerical approximation to Eq. (\ref{PF}) is then:
\BE
\mP(t_0,\tau)_{ij} \approx \frac{\textrm{number of particles
from box $i$ to box $j$}}{N_i} \ . \label{PFparticles}
\EE

Because of the time-dependence of the velocity field, the
results of the Lagrangian simulations will depend on both the
initial time $t_{0}$ and the duration of the simulation $\tau$.
Once these parameters are fixed, we can build a network
described by a transport matrix $\mP(t_0,\tau)$ that
characterizes the connections among each pair of nodes from
initial time $t_{0}$ to final time $t_{0} + \tau$. We interpret
$\mP(t_0,\tau)$ as the adjacency matrix of a weighted and
directed network, so that $\mP(t_0,\tau)_{ij}$ is the weight of
the link from node $i$ to node $j$.

The network constructed in this way characterizes the final
locations of all fluid elements a time $\tau$ after their
release at time $t_0$, but gives no information on particle
locations at intermediate times. Also, since each of the
matrices $\mP(t_0+k\tau,\tau)$, for $k=0,1,...,n-1$, is a
stochastic matrix, one can consider the discrete-time Markov
chain in which an initial vector giving occupation
probabilities $p(t_0)=\left(p_1(t_0),...,p_{N}(t_0)\right)$ for
the different boxes is evolved in time as
$p(t_n)=p(t_0)\mP(t_0,\tau)\mP(t_1,\tau)...\mP(t_{n-1},\tau)$,
where $t_k=t_0+k\tau$. This time evolution will not be exactly
equal to the true evolution $p(t_n)=p(t_0)\mP(t_0,n\tau)$, but
a Markovian approximation to it in which the memory of the
particle positions is lost after a time $\tau$. The Markovian
approximation may be reasonable in some circumstances and in
fact it has been successfully used in geophysical flow problems
\cite{Dellnitz2009,Froyland2012Three,Froyland2014}. In this
paper we do not assume any Markovian hypothesis and we work
with the full matrix $\mP(t_0,\tau)$ covering our time interval
of interest and describing only the initial and final states of
the transport process.

Despite not using any Markov assumption, replacing the
continuous flow system by a discrete network introduces
discretization errors. Even if the integration is done
accurately, the initial and final locations of the transported
particles are only specified up to a precision $\Delta$, given
by the linear side of the boxes. This implies that our network
approach does not display explicitly fluid structures smaller
than the box length-scale $\Delta$.

\section{The surface flow network of the Mediterranean sea}
\label{sec:Mediterranean}

We now apply the previous general procedures to build and
analyze the flow network associated to a realistic surface flow
in the Mediterranean sea.

The input velocity field originates from the Mediterranean
Forecasting System Model (physics reanalysis component). It is
a hydrodynamic model supplied by the Nucleus for European
Modelling of the Ocean (NEMO) which solves primitive equations
in spherical coordinates. NEMO has been implemented in the
Mediterranean at an horizontal resolution of
$\sfrac{1}{16}\,$degrees, and $72$ unevenly spaced vertical
levels \cite{Oddo2009}. It also slightly extends into the
Atlantic in order to resolve the Strait of Gibraltar. The model
uses vertical partial cells to fit the bottom depth shape. It
is forced by momentum, water and heat fluxes interactively
computed by bulk formulae using the $6$ hours,
$\sfrac{1}{4}\,$degree horizontal-resolution operational
analysis and forecast fields from the European Centre for
Medium-Range Weather Forecasts (ECMWF) while precipitation and
river runoffs are provided by monthly mean datasets. The
Dardanelles inflow is parameterized as a river and the
climatological net inflow rates are considered. Assimilated
data include sea level anomaly, remotely-sensed sea surface
temperature and in-situ temperature and salinity profiles.

We used daily horizontal velocity fields generated by the model
in the whole Mediterranean basin during $10$ years of
simulation ($2002-2011$) selecting only one layer at a nominal
depth of $7.9$ m. This layer extends in fact between 4.58 and
11.55 m depth, so that it has a vertical extension of 6.97 m.
For the integration time scales used here (values of $\tau$
always below three months) we can reasonably neglect motion to
other layers and consider only horizontal dynamics
\cite{dOvidio2004,Rossi2014}.


\subsection{Discretization}
\label{subsec:Mediterranean:discretization}

\begin{figure}
\includegraphics[width=\columnwidth,clip=false]{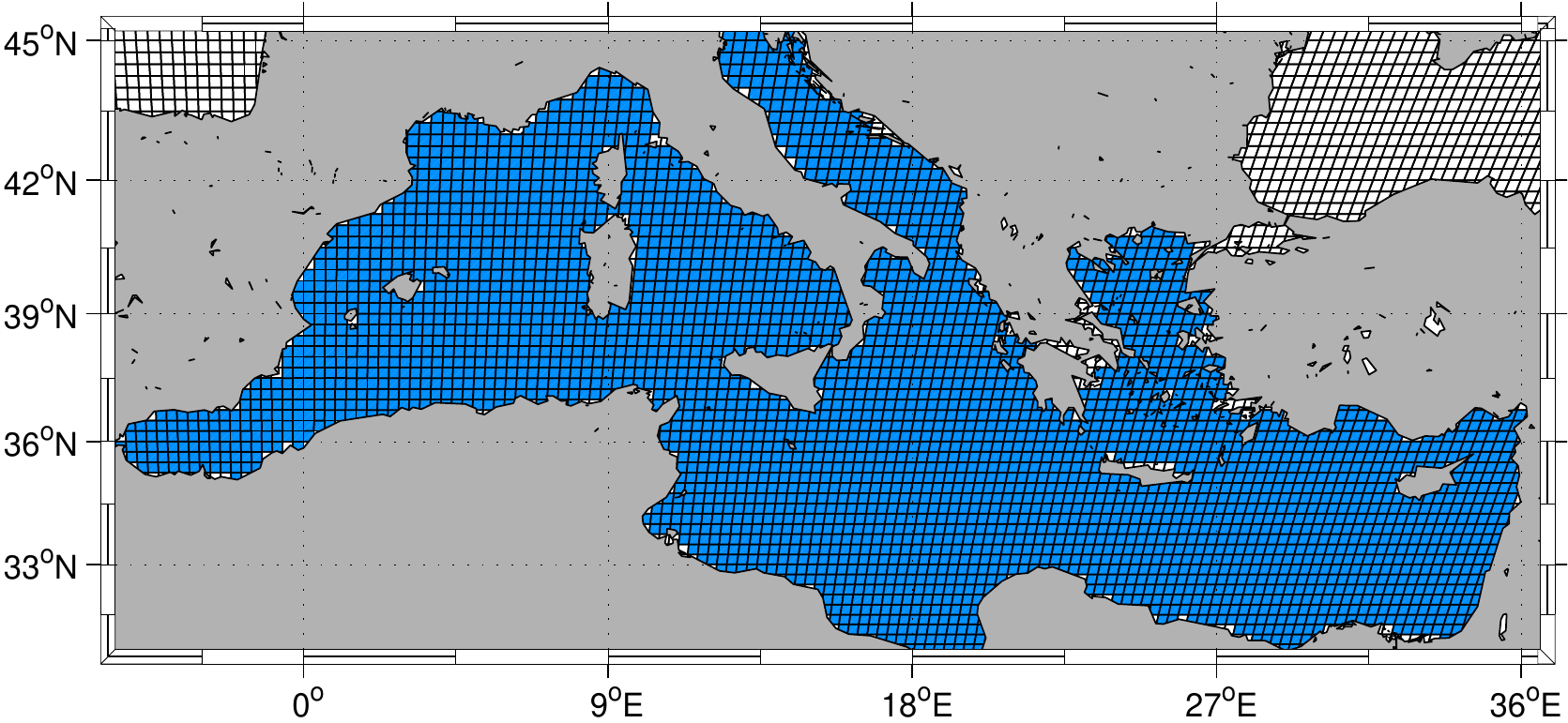}
\caption{Discretization of the Mediterranean sea (blue region)
into $N=3270$ equal-area boxes $\{B_i, i=1,...,N\}$.}
\vspace*{0.0cm} \label{fig:discrgrid}
\end{figure}

To switch from continuous space to discrete nodes we partition
the above-described horizontal near-surface Mediterranean layer
into $3270$ two-dimensional square boxes. We imposed the
equal-area constraint defining the cells in a sinusoidal
projection given by coordinates $x$ and $y$ related to the
standard longitude $\varphi$ and latitude $\phi$ by
\begin{equation}
 x =\varphi \cos \phi \qquad;\qquad y =\phi
\end{equation}
In these $x,y$ coordinates, boxes are squares of side $0.25$
degrees, or $\Delta=27.78~km$ (see Fig.~\ref{fig:discrgrid}).
The area $\Delta^2$ of each box is $771.9~km^2$. The ``amount
of water in a box $B_i$" is then related to its area $\Delta^2$
through a simple multiplication by the layer thickness (6.97
m), returning a value of $5.38\times 10^9~m^3$ per box.

The resolution of the model-generated velocity field is much
finer than the discretization we use for network construction.
In this sense the dynamics represented in the flow network is a
coarse-graining of the simulated Mediterranean flow, keeping
the effect of the small scales only in a statistical sense. The
most energetic features of the Mediterranean flow are mesoscale
structures \cite{Millot2005} ranging from 10 km to a few
hundred km. With the value of $\Delta$ we use, our network
description displays most of the mesoscale range, and neglects
submesoscales, which anyway are only marginally resolved by the
NEMO implementation.

\subsection{Lagrangian simulation}
\label{subsec:Mediterranean:Lagrangian}

To characterize the transport phenomena, $N_i=500$ ideal fluid
particles were released in each box $B_i$, providing enough
statistics to estimate $\mP(t_0,\tau)$. We simulated the motion
of these $3270\times 500=1.635\times 10^6$ particles by
integrating the trajectories in the velocity field using a
fourth-order Runge-Kutta algorithm. The velocity at any
arbitrary point in the sea is computed with a bilinear
interpolation from the input data. We used a time step of $1$
day (the same resolution as the data). We also tested shorter
intervals using a cubic interpolation but no significant
improvement was found. The two key-parameters of the
simulations are the starting time $t_{0}$ and the tracking time
$\tau$.


\subsection{Network construction}
\label{subsec:Mediterranean:adjacency}

The simulation provides the initial and final positions for
each particle, allowing us to compute the transport matrix
$\mP(t_0,\tau)$ from Eq. (\ref{PFparticles}). A directed link
is established from node $i$ to node $j$ if and only if
$\mP(t_0,\tau)_{ij}$ is non-vanishing. In that case its value
gives the weight of such a link. Due to numerical limitations,
some trajectories end up prematurely by ``beaching" onto land
areas outside of the partition $\{B_i\}$. Then, the denominator
$N_i$ in Eq. (\ref{PFparticles}) is taken as the number of
particles still in the sea at the end of the integration time
$\tau$. Since the beaching effect is small, affecting less than
5\% of all particles in the longest simulations presented here
(and only for near-shore boxes), we still assume in the
following that the convenient equal-area condition remains
approximately valid.

Note that the Lagrangian integration is done under the full
resolution of the velocity field ($\sfrac{1}{16}^{\circ}$).
This means that particle trajectories contain the small-scale
features produced by the model during time $\tau$. While such
details are not explicitly present in the network description
$\mP(t_0,\tau)$ after coarse-graining the initial and final
positions to the box size $\Delta$, their effects have been
incorporated in a statistical way.


\section{Network properties}
\label{sec:network}

We now interpret the transport matrices $\mP(t_0,\tau)$, for
several values of $t_0$ and $\tau$, as the adjacency matrices
of directed and weighted flow networks. We can calculate for
them all the standard quantities characterizing the topology of
networks, such as degree, clustering, betweenness,
etc.\cite{Newman2010}. But following the aim stated in the
Introduction, we will concentrate here in network quantities
that can give insight in (horizontal) dispersion and mixing
processes, and in the identification of coherent regions.

\subsection{Dispersion and mixing}
\label{subsec:network:dispersion}

Important properties of geophysical flows depend on their
dispersion characteristics, i.e. how far away can the fluid be
transported during some time, and how diverse are the target
regions. Mixing of fluid with different characteristics,
another process of great geophysical relevance, will occur at a
particular place if fluid from different origins arrives there
at a particular time.

In dynamical systems approaches to flow processes, a standard
way to quantify dispersion is by means of the finite-time
Lyapunov exponent (FTLE). It is defined as\cite{Shadden2005}
\BE
\lambda(\bx_0,t_0,\tau)=\frac{1}{2|\tau|}\log \Lambda_{max}
\label{FTLE}
\EE
where $\Lambda_{max}$ is the maximum eigenvalue of the
Cauchy–Green strain tensor:
\BE
C(\bx_0,t_0,\tau) = \left(\nabla
\Phi_{t_0}^\tau(\bx_0)\right)^T \nabla \Phi_{t_0}^\tau(\bx_0)
\EE
constructed from the Jacobian matrix $\nabla
\Phi_{t_0}^\tau(\bx_0)$ of the flow map. $M^T$ means the
transpose of the matrix $M$. For $\tau>0$ this is the forward
FTLE. By running the flow map backwards in time ($\tau<0$) we
get the backwards FTLE field, which quantifies the strength of
mixing into a particular location. The interpretation of
(\ref{FTLE}) is that an initial circle of infinitesimal
diameter $\delta$ located at $\bx_0$ at $t_0$ will become an
ellipse of major axis $e^{\tau \lambda(\bx_0,t_0,\tau)}\delta$
after being advected by the flow during a time $\tau$. The
minor axis will be a decreasing function of $\tau$, contracting
at an exponential rate related to a negative Lyapunov exponent
that can be computed from the second eigenvalue of
$C(\bx_0,t_0,\tau)$.

\begin{figure}
  \includegraphics[width=\columnwidth,clip=true]{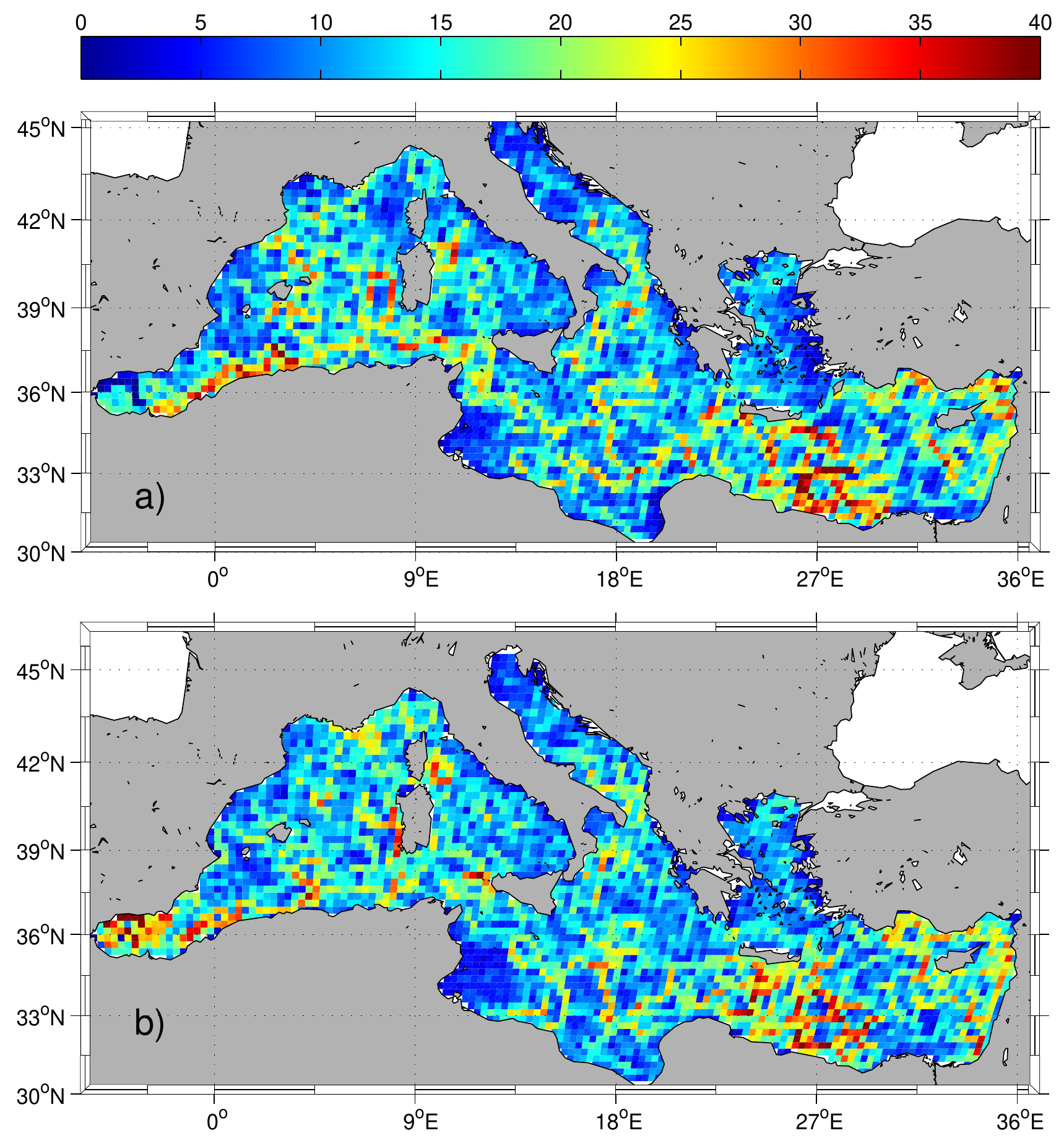}
  \caption{\label{fig:deg} Degree of the nodes in the flow network defined by
  $\mP(t_0,\tau)$, for $t_0$=July 1st 2011 and $\tau=15$ days. a) The in-degree $K_I(i)$.
  b) The out-degree $K_O(i)$.}
\end{figure}

An obvious quantity in the network interpretation suitable to
be related to dispersion and mixing is the {\sl degree} of a
node. Since our network is directed, we should distinguish
between the in-degree $K_I(i)$, i.e. the number of links
pointing to a particular node $i$, and the out-degree $K_O(i)$,
the number of links pointing out of it. Figure \ref{fig:deg}
displays these quantities at the geographical locations defined
by the nodes of the Mediterranean network for particular values
of $t_0$ and $\tau$. High values of the degrees appear
associated to the strong and unstable currents in the southern
part of the basin\cite{Millot2005}. Low degree values are
observed in regions where the circulation is rather slow, such
as the Tunisian continental shelf and the semi-enclosed seas
(e.g. Adriatic and Aegean). Generally, the values of the in-
and out degree- tend to increase with $\tau$. With respect to
the dependence on $t_0$, degree values tend to be slightly
higher in winter than in summer.

\begin{figure}
  \includegraphics[width=\columnwidth]{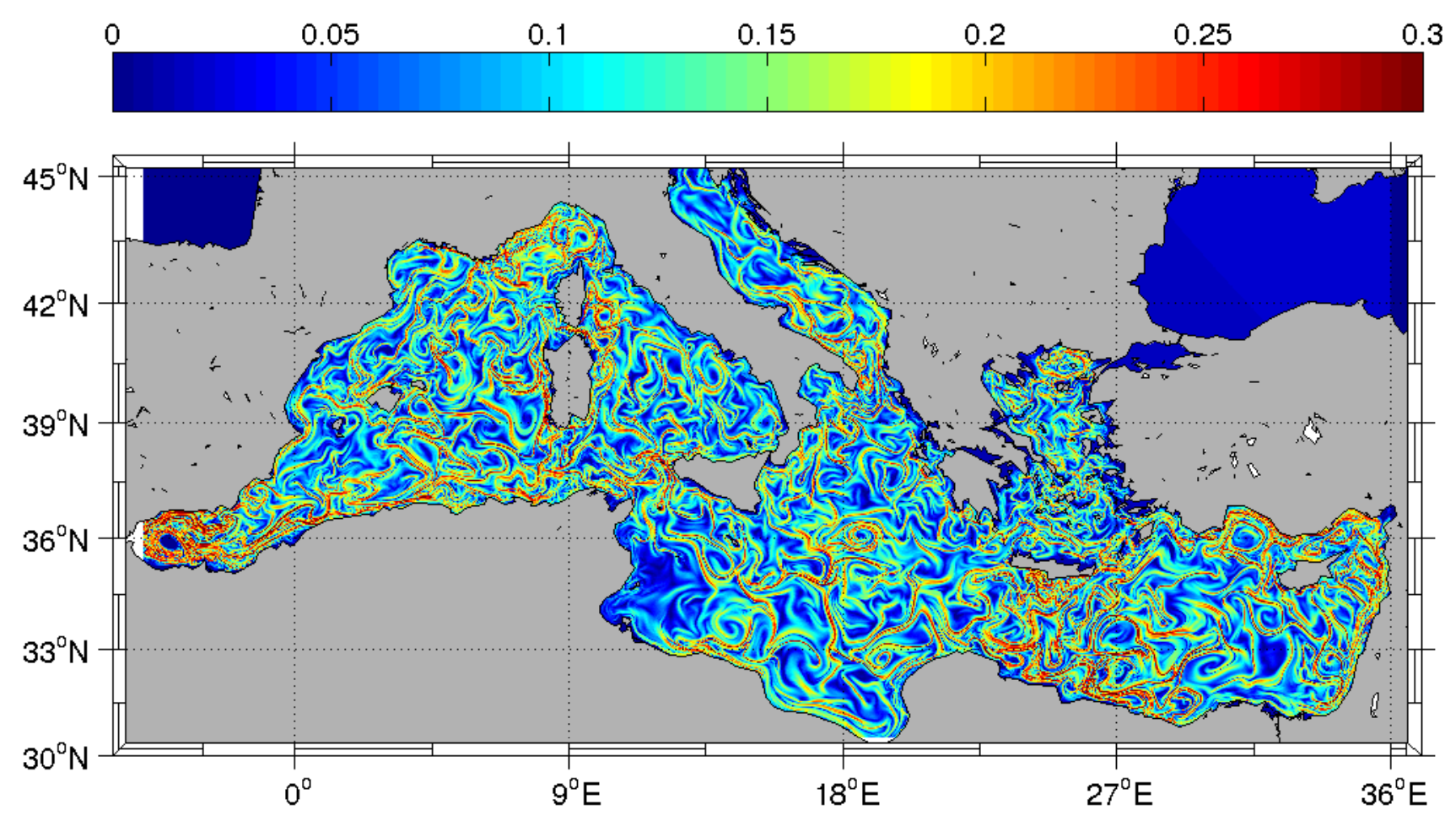}
  \caption{An example of forward FTLE field $\lambda(\bx_0,t_0,\tau)$
  at $t_0=$July 1st 2011, and $\tau=15$
days. Color bar in day$^{-1}$}
  \label{fig:FTLE}
\end{figure}

\begin{figure}
  \includegraphics[width=\columnwidth]{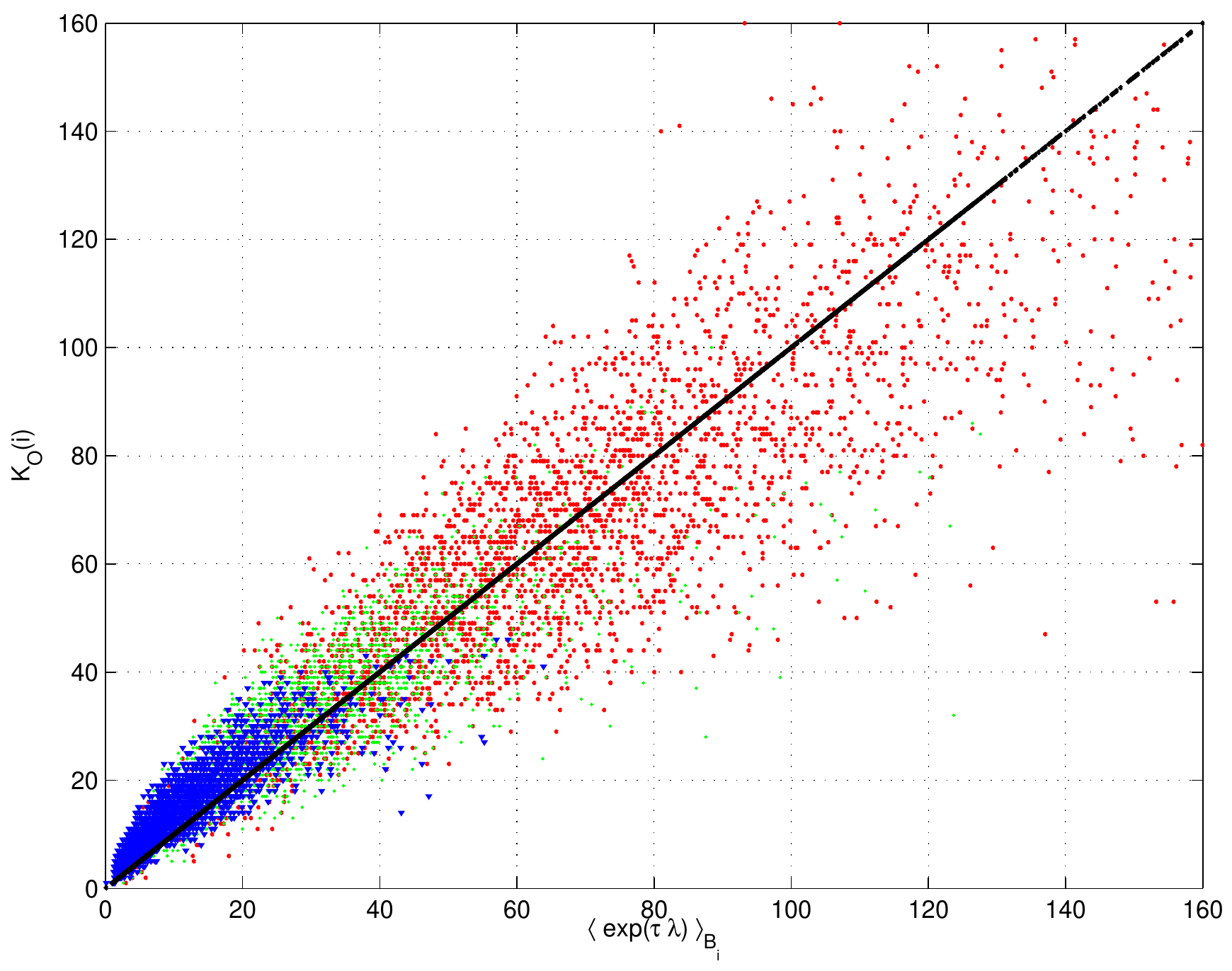}
  \caption{Values of the out-degree $K_O(i)$ of each node $i$ vs the average value of
  the stretching factor $e^{\tau\lambda}$ in that node. $t_0=$July 1st 2011.
  Blue symbols are from $\tau=15$ days, green from $\tau=30$ days and red from
  $\tau=60$ days. Black line is the main diagonal.}
  \label{fig:KOivsStretch}
\end{figure}

A first problem in relating these network properties to the
actual physics of dispersion and mixing is that their values
are dependent on the spatial scales chosen for discretization
(there is also a dependency on the numbers $N_i$ of particles
used to compute the transport matrices, but it disappears for
large $N_i$). This problem is easy to solve by recalling that
every box has an associated area. Dealing first with the
out-degree case for definiteness, $K_O(i)$ is proportional to
the total area of all nodes that received some contents from
the initial node $i$. This quantity has a well-defined meaning
that can be related to the continuous flow dynamics with only a
minor dependence on the discretization procedure. Since here
all boxes have the same area $\Delta^2$, the area corresponding
to the out-degree of node $i$ is $K_O(i)\Delta^2$. We can use
generic ideas of chaotic dynamics to obtain heuristically a
more precise relationship between two quantifiers of
dispersion: the degree and the Lyapunov exponent. In regions
dominated by hyperbolic structures, each of the fluid boxes
will be stretched into a long and thin filament after a
sufficiently long time $\tau$ (see Appendix A). If we want to
compute the number of boxes reached by it, it is enough to
consider its length, since the width quickly becomes smaller
than the box size $\Delta$. Let us consider an initial line of
length $L(t_0)\approx \Delta$ inside the initial box $B_i$. A
small segment of it, of length $dl(t_0)$ at position $\bx_0 \in
B_i$ will become elongated by a factor given by the local FTLE:
$dl(t_0+\tau)=dl(t_0)e^{\tau \lambda(\bx_0,t_0,\tau)}$.
Integrating over the initial positions along the line we get an
estimation of the final length $L(t_0+\tau)$ of the filament. A
better estimation $\bar L(t_0+\tau)$ of this length can be done
by averaging over positions transverse to the line, to take
into account different locations of the initial line in the
box:
\BE
\bar L(t_0+\tau) \approx \frac{1}{\Delta} \int_{B_i} d\bx_0 e^{\tau
\lambda(\bx_0,t_0,\tau)} \ ,
\EE
where the longitudinal and transverse integrations have been
combined into the integration of $\bx_0$ over the area $B_i$.
The area of the boxes covered by the filament is ${\cal
A}(t_0+\tau) \approx \bar L(t_0+\tau)\Delta$ so that the
out-degree of the initial box will be
\BA
K_O(i) = \frac{{\cal A}(t_0+\tau)}{\Delta^2} &\approx& \nonumber \\
   \frac{1}{\Delta^2} \int_{B_i} d\bx_0 e^{\tau
   \lambda(\bx_0,t_0,\tau)} &\equiv& \left<
   e^{\tau\lambda(\bx_0,t_0,\tau)}\right>_{B_i}.
\label{KOi}
\EA
Thus, we have a useful relationship between a natural quantity
in the network description of fluid flows and a standard
characterization of dispersion in the dynamical systems
approach to such flows: the degree of a node associated to a
box is the average or coarse-graining of the stretching factor
$e^{\tau\lambda}$ in that box. We can check the validity of the
above heuristic arguments by comparing directly the values of
$K_O(i)$ obtained from our flow network and the right-hand-side
of (\ref{KOi}). Figure \ref{fig:FTLE} shows an example of FTLE
field obtained at time $t_0=$ July 1st 2011, and $\tau=15$
days. Figure \ref{fig:KOivsStretch} shows the clear correlation
between the two quantities. Three values of $\tau$ are plotted
to appreciate the general validity of the relationship. We
attribute the deviations with respect to the exact identity to
the fact that the filament-type arguments are only valid for
sufficiently large $\tau$ and in regions dominated by strain.
Also, our arguments neglect the presence of filament foldings
that sometimes would occupy the same box, and of associated
saturation effects. In addition quantization effects arising
from the discrete nature of $K_O$ are visible at small degree
values.

Expression (\ref{KOi}) suggests defining
\BE
H_i^0(t_0,\tau) \equiv \frac{1}{\tau} \log K_O(i)
\label{Hi0}
\EE
so that
\BE
\left< e^{\tau\lambda(\bx_0,t_0,\tau)}\right>_{B_i} = e^{\tau
H_i^0(t_0,\tau)} \ . \label{expHi0}
\EE
From the convexity of the exponential function, we have
$H_i^0(t_0,\tau) \ge \left<
\lambda(\bx_0,t_0,\tau)\right>_{B_i}$. The previous expressions
are reminiscent of the properties of the topological entropy of
a dynamical system, as giving the exponential growth in time of
the length of a material line\cite{Tel2006}. Pushing forward
the analogy, we can define a sequence of R\'{e}nyi-like
entropies\cite{Renyi1970} associated to a particular node $i$:
\BE
H_i^q(t_0,\tau) \equiv \frac{1}{ (1-q)|\tau|}\log \sum_{j=1}^N
\left( \mP(t_0,\tau)_{ij}\right)^q \ , \label{Hiq}
\EE
which we call {\sl network entropies}. Due to their dependence
on the finite-size of the partition, they are related to the
$\epsilon$-entropies discussed by \citet{Boffetta2002}. Note
however that here the transport matrix involves only two states
of the trajectories, separated by an interval of time $\tau$
which remains finite, and the dependence on the initial
location, box $B_i$, is kept. The entropies $H_i^0$ and $H_i^1$
should be understood as defined by the limits $q\rightarrow 0$
and $q\rightarrow 1$, respectively. All the network entropies
measure the {\sl diversity} in the amounts of fluid received by
the nodes connected to a given box, but weighting them in
different ways: In $H_i^0$ all nodes are counted equally
independently of the amount of water they receive, so that it
informs only about the degree as seen in Eq. (\ref{Hi0}); for
increasing values of $q$ nodes receiving more water are
weighted with increasing strength. Although the network
entropies have been introduced here in the particular context
of flow networks, we note that they can be defined for any
weighted network, giving generalizations of the degree to
quantify the unevenness of the weight distribution towards the
nodes connected to a given one.

\begin{figure}
  \includegraphics[width=\columnwidth,clip=true]{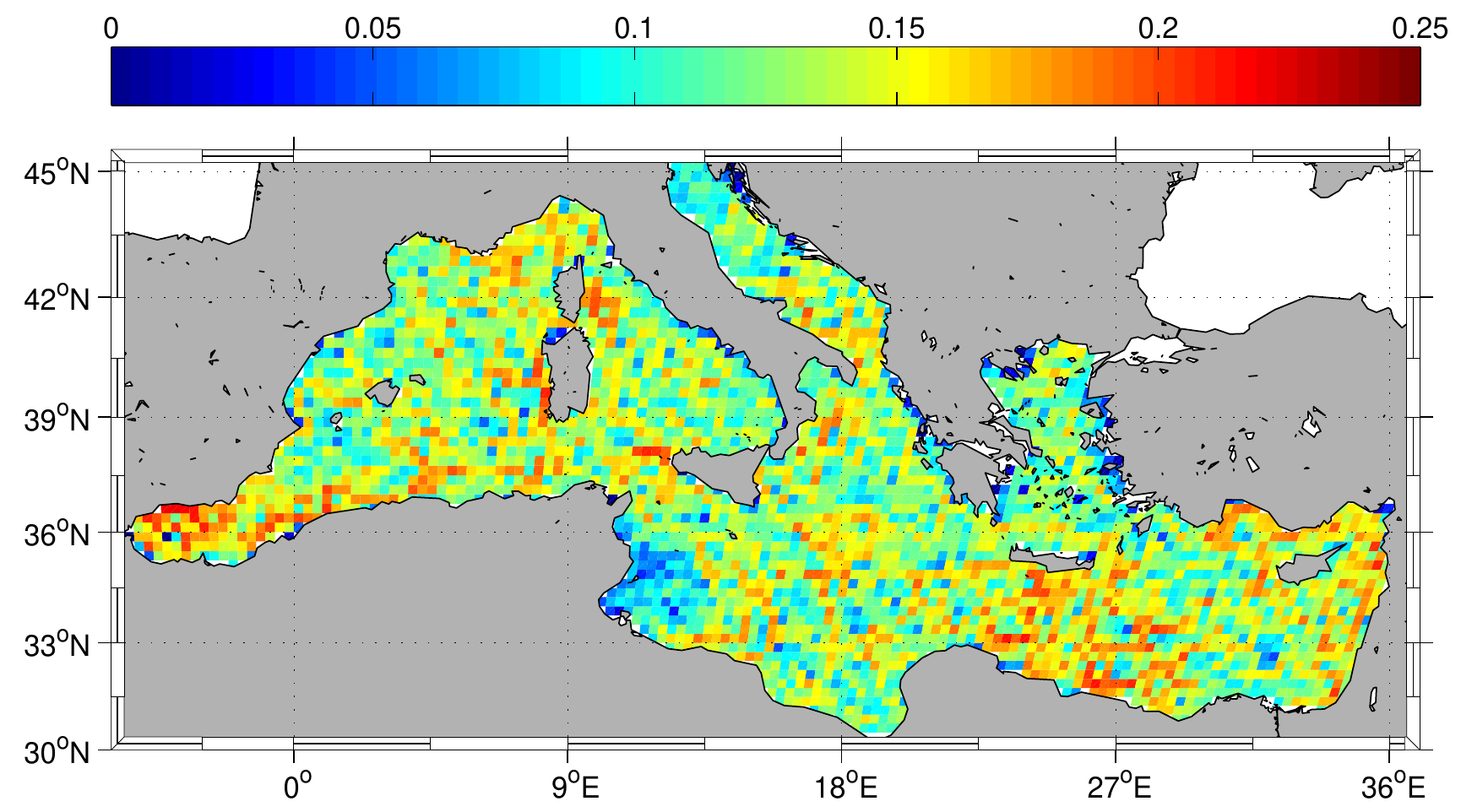}
  \caption{The network entropy $H_i^1(t_0,\tau)$, for $t_0=$July 1st 2011, and $\tau=15$
days. Color bar in day$^{-1}$.}
  \label{fig:FTE}
\end{figure}

Applying l'H\^{o}pital's rule to the definition of the network
entropy of order $q=1$ one gets:
\BE
H_i^1(t_0,\tau)=-\frac{1}{\tau} \sum_{j=1}^N \mP(t_0,\tau)_{ij}
\log \mP(t_0,\tau)_{ij} \ .
\EE
It gives the amount of information (per unit of time) gained by
observing the position of a particle at time $t_0+\tau$,
knowing that it was initially (time $t_0$) somewhere in box
$B_i$. This quantity is precisely the discrete finite-time
entropy studied by \citet{Froyland2012Finite}. Figure
\ref{fig:FTE} shows its spatial distribution in the
Mediterranean sea for particular values of $t_0$ and $\tau$.

\begin{figure}
  \includegraphics[width=\columnwidth,clip=true]{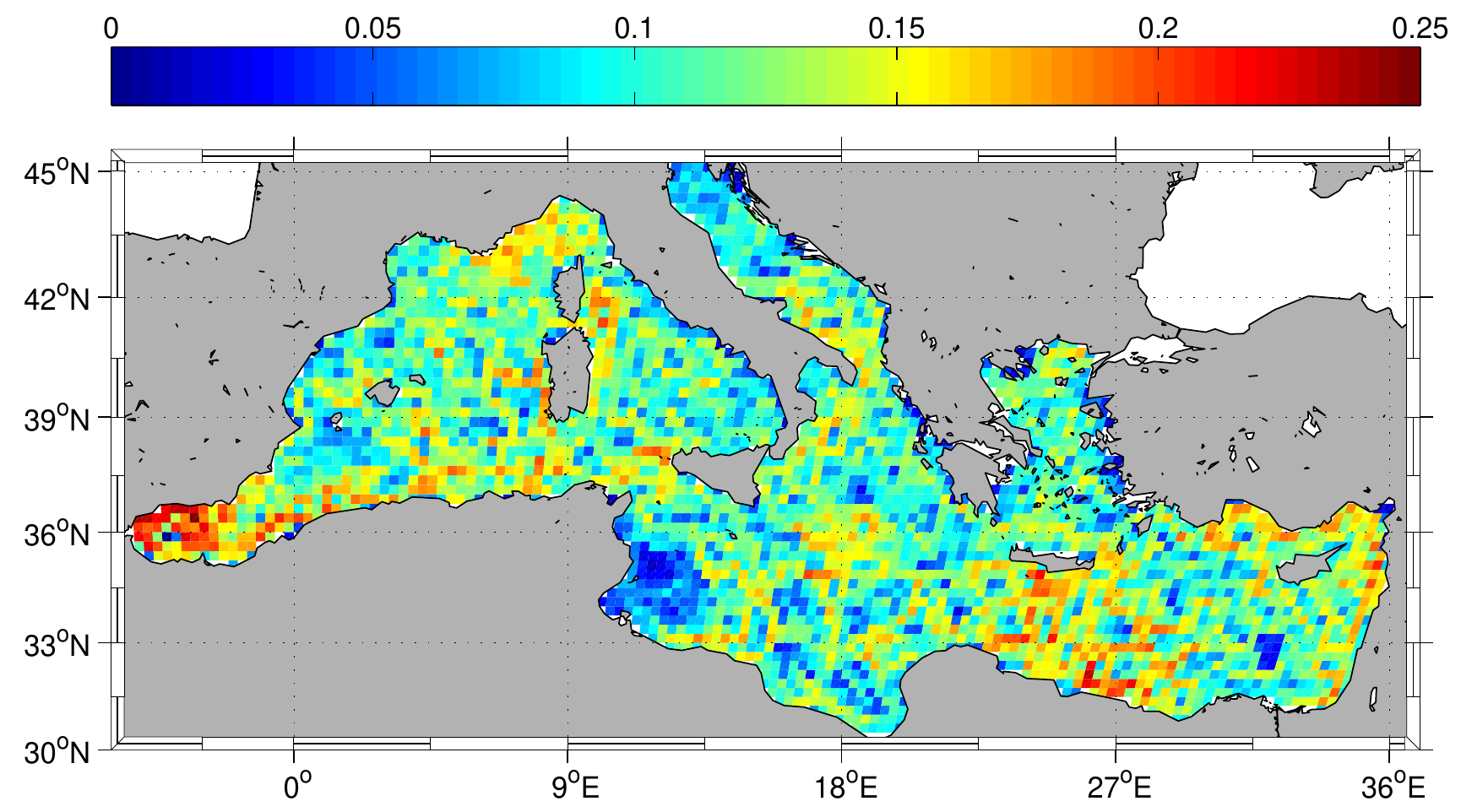}
  \caption{Coarse-graining
  of the Lyapunov field in Fig. \ref{fig:FTLE} into the discretization boxes:
  $\lambda_i(t_0,\tau)\equiv \left<\lambda(\bx_0,t_0,\tau)\right>_{B_i}$.
  $t_0=$July 1st 2011, and $\tau=15$. Color bar in day$^{-1}$.}
  \label{fig:discreteFTLE}
\end{figure}

\begin{figure}
  \includegraphics[width=\columnwidth]{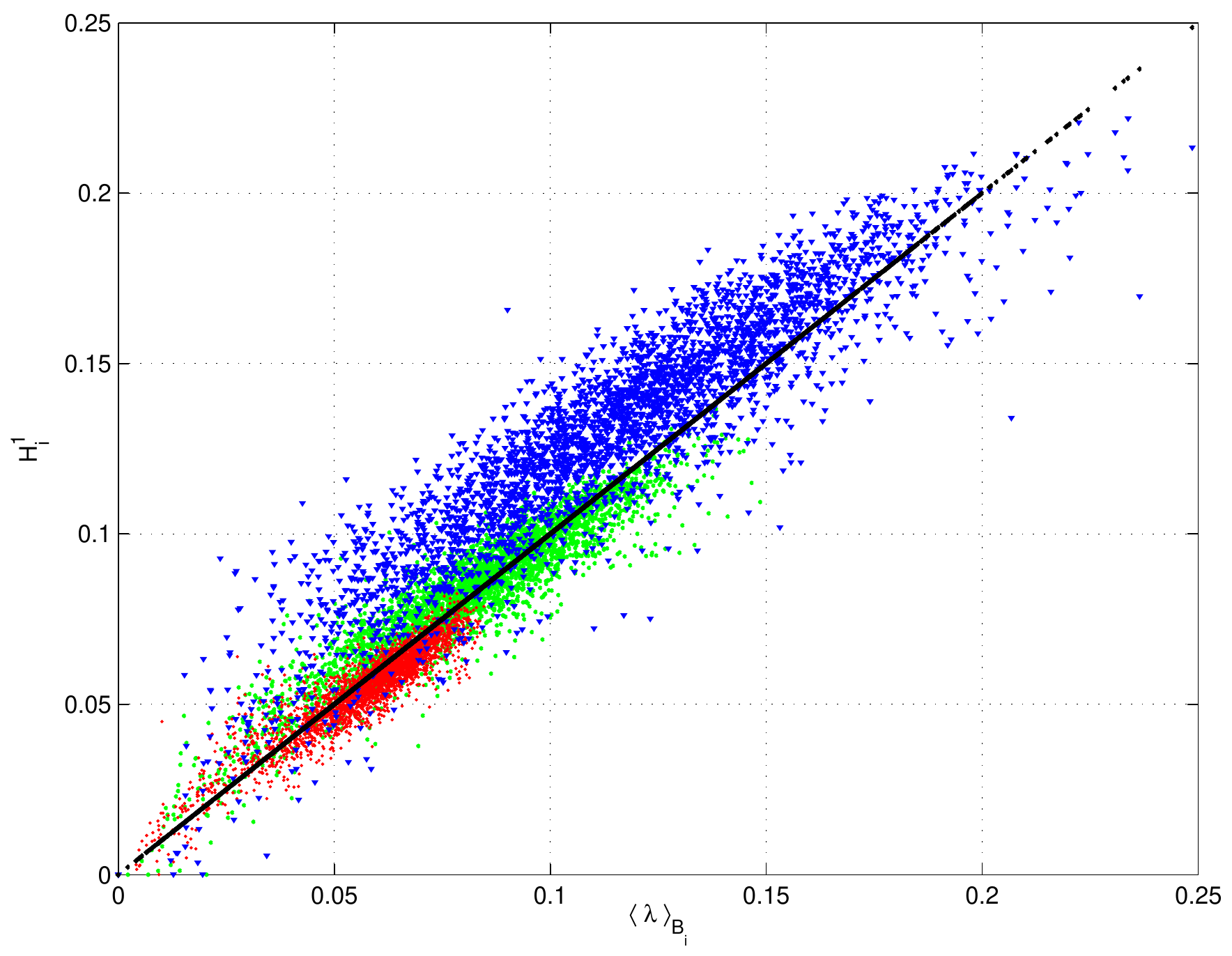}
  \caption{Values of the network entropy $H_i^1(t_0,\tau)$ of each node $i$ vs the average value of
  the Lyapunov exponent in that node, $\lambda_i(t_0,\tau)$.  $t_0=$July 1st 2011.
  Blue symbols are from $\tau=15$ days, green from $\tau=30$ days and red from
  $\tau=60$ days. Black line is the main diagonal.}
  \label{fig:FTEvsFTLE}
\end{figure}

The standard Pesin-like results relating the metric or
Kolmogorov-Sinai entropy to the sum of positive Lyapunov
exponents \cite{Boffetta2002,Castiglione2010,Cencini2010}
suggest that, at least for large $\tau$, the entropy $H_i^1$
would give a good approximation to the values of the FTLE field
averaged over each box $B_i$: $\lambda_i(t_0,\tau)\equiv
\left<\lambda(\bx_0,t_0,\tau)\right>_{B_i} \approx
H_i^1(t_0,\tau)$. Appendix A gives calculations supporting this
claim in an heuristic way. Figure \ref{fig:discreteFTLE} shows
the geographical distribution of $\lambda_i(t_0,\tau)$ and Fig.
\ref{fig:FTEvsFTLE} compares both quantities for several values
of $\tau$. The entropies tend to be slightly larger than the
Lyapunov exponents for $\tau = 15$ days, but both quantities
approach each other and become well correlated for larger
$\tau$.

\begin{figure}
  \includegraphics[width=\columnwidth]{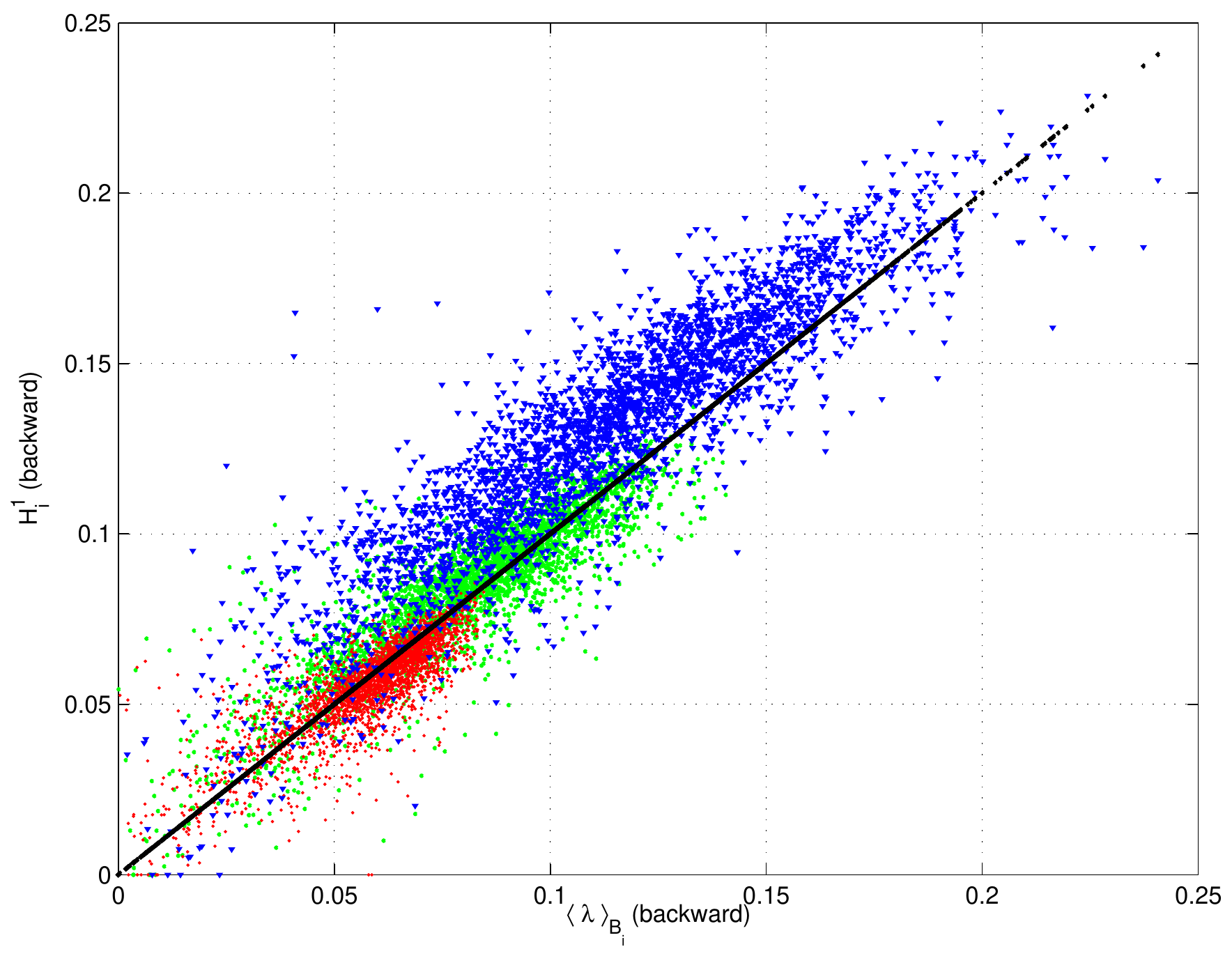}
  \caption{Values of the network entropy $H_i^1$ at each node $i$, computed from the
  backwards-dynamics network given by $\mP(t_0+\tau,-\tau)$ (Eq. (\ref{Pback})),
  vs the average value of
  the backwards Lyapunov exponent in that node, $\lambda_i(t_0+\tau,-\tau)$.  $t_0=$July 1st 2011.
  Blue symbols are from $\tau=15$ days, green from $\tau=30$ days and red from
  $\tau=60$ days. Black line is the main diagonal.}
  \label{fig:backFTEvsFTLE}
\end{figure}

For definiteness we have been discussing quantities related to
the forward time evolution: out-degree, forward Lyapunov
exponents, etc. The network entropies can also be defined for
the backward time evolution. Construction of the
backwards-dynamics network can be achieved by redoing the
launching of particles and running the Lagrangian integration
for negative time, or much simpler, by
recognizing\cite{Froyland2012Finite} that the backward
evolution is given by the matrix
\BE
\mP(t_0+\tau,-\tau)_{ij}=\frac{\mP(t_0,\tau)_{ji}}{\sum_{k=1}^N
\mP(t_0,\tau)_{jk}} \ . \label{Pback}
\EE
The network entropies in Eq. (\ref{Hiq}) can now be directly
computed for the backward flow network defined by
$\mP(t_0+\tau,-\tau)$, and they will be related to backwards
Lyapunov fields, which give a measure of mixing of fluid coming
from different origins. As an example we show in Fig.
\ref{fig:backFTEvsFTLE} the relationship between the backwards
entropy $H_i^1(t_0+\tau,-\tau)$ and the coarse-grained
backwards Lyapunov exponent $\lambda_i(t_0+\tau,-\tau)$. Again
both quantities are similar for sufficiently large $\tau$ and
the same qualitative features as in Fig. \ref{fig:FTEvsFTLE}
are observed.

Summarizing this Section, we have defined a family of
entropy-like quantities completely in terms of the transport
matrix characterization of the flow network. At least two of
them, $H_i^0$ and $H_i^1$, are related to standard dispersion
and mixing quantifiers in the description of fluid flows. The
higher order entropies $H_i^q$ are related to the generalized
Lyapunov exponents \cite{Boffetta2002,Cencini2010}
characterizing successive moments of the Lyapunov field, as
discussed in Appendix A. We do not claim that these
relationships are exact for finite values of $\tau$ and
$\Delta$. Instead, we find numerical deviations from them
(Figs. \ref{fig:KOivsStretch}, \ref{fig:FTEvsFTLE} and
\ref{fig:backFTEvsFTLE}) which decrease for increasing $\tau$.
We expect the same to happen when decreasing $\Delta$. The
important point is that, once the network matrix
$\mP(t_0,\tau)$ has been constructed, the entropies in Eq.
(\ref{Hiq}) provide a computationally very cheap way to assess
quantities of geophysical interest such as local dispersion,
stretching and mixing. In fact the simplest network quantifiers
such as the in- and out-degrees are already suitable for that,
being related to $H_i^0$. The qualitative information displayed
in figures such as \ref{fig:deg}b and \ref{fig:FTE} or
\ref{fig:discreteFTLE} is very similar. Also, even if we should
have $H_i^0\ge H_i^1 \approx\lambda_i$, in our examples the
numerical values of $H_i^0$ are only slightly larger than those
of $H_i^1$. We have to mention that we have been working under
the hypothesis of boxes $\{B_i\}$ of equal areas. Expression
(\ref{Hiq}) would need corrections in a more general case. See
for example the case of $H_i^1$ in \citet{Froyland2012Finite}.

\subsection{Identification of coherent regions}
\label{subsec:network:coherent}

\subsubsection{Coherent regions as network communities}
\label{CoherenceAsCommunities}

Most work in the dynamical systems approach to fluid transport
aims at identifying ``barriers to transport" locating the
borders of regions that do not exchange much fluid among them.
The set-oriented approach focusses on the regions themselves
rather than on the borders. Almost-invariant fluid regions have
been defined as regions of the fluid domain remaining
relatively isolated (according to a suitably defined metrics)
from the rest of the fluid\cite{Froyland2003,Froyland2005}. In
generic time-dependent flows these regions will not be fixed in
space but they will be transported by the mean flow, and the
concept of coherent pairs, relating initial and final set
positions has been developed
\cite{Froyland2010,Santitissadeekorn2010,Froyland2012Three}.
Formulating this problem in the context of network theory would
require building on techniques for bipartite graphs. In our
present case study, the global flow in the Mediterranean sea,
land masses play an important role in restricting the flow, so
that coherent regions that remain fixed with respect to the
coasts are the most relevant ones for many applications. In
particular, when considering environmental conservation
strategies and marine reserves
\cite{Nilsson-Jacobi2012,Thomas2014,Rossi2014}, one looks for
the connectivity among marine zones, or
provinces\cite{Rossi2014}, occupying localized regions of the
sea. Thus we focus here on finding a partition of the sea into
{\sl self-coherent}, or {\sl almost-invariant} regions,
associated to relatively stable circulation patterns, from the
point of view of network theory. We want these regions to be
well-mixed internally, and with little interchange with the
exterior. In the language of networks this translates to
partitioning the network into subgraphs with high internal
connectivity, and small connectivity among them. This is the
standard problem of community detection in
networks\cite{Newman2010,Danon2005,Lancichinetti2009,Fortunato2010,Aldecoa2013},
for which many different and powerful techniques are available.
In fact, most of the approaches used so far to partition fluid
motion into almost-invariant
sets\cite{Froyland2003,Froyland2005,Speetjens2013} employ
classical spectral techniques for graph
partition\cite{Newman2010}, which use the eigenvectors or
singular vectors of the transport matrix (or other matrices
derived from it). We note that the methodologies in Refs.
\cite{Froyland2003,Froyland2005} find almost-invariant sets in
the sense that loss and gain of fluid is minimized. But the
condition of strong internal mixing, which we consider
important in geophysical applications, is not imposed.

Here we address the community detection problem with a
state-of-art network-theory approach, the \textit{Infomap}
algorithm \cite{Rosvall2008}. The method is based on the
probability flow of random walks in the network moving with
transition probabilities given by the adjacency matrix
$\mP(t_0,\tau)$, and on exploiting the properties of
information compression in the description of that probability
flow. \textit{Infomap} finds the partition of the network
minimizing the average size of the codeword needed to describe
inter- and intracommunity transitions. A succinct description
of the method is provided in Appendix B. We believe this
methodology is specially suited to partition flow networks for
the following reasons: First, it takes into account the
``direction" and ``weight" of each link, important
characteristics defining our flow network. The standard
spectral methods and most modularity-optimization algorithms
take as input a symmetrized version of the network. Second,
\textit{Infomap} does not require to fix \textsl{a priory} the
number of communities forming the domain partition. Third,
\textit{Infomap} does not impose similar sizes to the
communities so that it does not suffer from the ``resolution
limit" \citep{Fortunato2007} restricting the minimum community
size detectable by most algorithms, including spectral methods.
This is important in geophysical flow networks since ocean
structures of different sizes coexist in the sea, some of them
arising from geographical accidents, bathymetry, etc.

The method has also some limitations. One of them is the
``field of view limit"\cite{Schaub2012} due to the use of a
single-step transition matrix $\mP(t_0,\tau)$. In general this
imposes that the detected communities are only those with
intense intracommunity connections (clique-like). For our
application this feature may become convenient since
\textit{Infomap} will identify as communities only regions well
mixed internally by the flow.

Since \textit{Infomap} consider random walkers exploring the
network with the transition probabilities in the matrix
$\mP(t_0,\tau)$, one is tempted to confuse these walkers with
the Lagrangian particles advected by the flow. But this is not
correct. $\mP(t_0,\tau)$ contains relationships between initial
and final positions of particles after a time $\tau$, but does
not describe in detail the trajectories at intermediate times.
In addition it can not be used beyond that time since in
time-dependent velocity fields flow connectivity will change
with the initial time $t_0$, defining the dynamic network.
\textit{Infomap} unveils the graph structures present in the
single matrix $\mP(t_0,\tau)$ by releasing random walkers that
evolve in a virtual time not directly related to the physical
time.

Hydrodynamical provinces delimited by \textit{Infomap} in the
Mediterranean surface flow were already studied by
\citet{Rossi2014}, who discussed also their implications for
the design of marine reserves. Here we concentrate in the
technical aspects and compare with alternative methods.


\subsubsection{Quality parameters}
\label{QualityParameters}

A standard way to asses the quality of a network partition is
by computing a modularity
parameter\cite{Newman2004Finding,Newman2010}. But this involves
comparison with a random null model than in the case of flow
networks has no obvious meaning. Then we prefer to use
alternative quantifiers with a direct interpretation in terms
of fluid connectivity. Here we define a {\sl coherence ratio}
and a {\sl mixing parameter}.

If coherent regions $A$ are understood as almost-invariant
areas of fluid, this means that they are mapped by the flow
nearly into themselves after a time $\tau$:
\begin{equation}
\Phi_{t_0}^\tau(A) \approx A \ .
\end{equation}
To measure how well this is achieved one can introduce the {\sl
coherence ratio}\cite{Froyland2003,Froyland2005}:
\begin{equation}
\rho_{t_0}^\tau(A) = \frac{m(A \cap \Phi_{t_0+\tau}^{-\tau} (A))}{m(A)}
\end{equation}
where, as before, $m(C)$ is the area of set $C$, but it can be
generalized to other measures. We have $\rho_{t_0}^\tau(A)\le
1$ and values close to unity indicate that $A$ is a truly
almost-invariant set.

In our discrete set-up, we consider sets $A$ made of our boxes
$\{B_i,i=1,...,N\}$: $A=\cup_{i \in \mathcal{I}} B_i$, where
$\mathcal{I}$ is the set of indices identifying the boxes $B_i$
making $A$. The coherence ratio is
now\cite{Froyland2003,Froyland2005}
\BE
\rho_{t_0}^\tau(A) = \frac{\sum_{i,j \in \mathcal{I}} m(B_{i})
\mP(t_0,\tau)_{ij}}{\sum_{i \in \mathcal{I}} m(B_{i})} \ .
\label{coherence}
\EE

For a partition of the fluid domain into $p$ communities or
provinces: ${\cal P}=\{A_1,...,A_p\}$, a global quality figure
of the partition is
\BE
\rho_{t_0}^\tau({\cal P}) \equiv \frac{1}{p} \sum_{k=1}^p
\rho_{t_0}^\tau(A_k) \ ,
\label{globalcoherence}
\EE
where again a good partition would be indicated by a value
close to 1. When communities are of very different sizes it may
be appropriate to weight the average in Eq.
(\ref{globalcoherence}) with these sizes, but we keep the
present definition to allow comparison with previous works.

Physically we can say that $\rho_{t_0}^\tau({\cal P})$
represents the fraction of tracers that at time $t_{0} + \tau$
are found in the same province where they were released at time
$t_{0}$. The definition involves the initial and final
positions, but gives no information on the particle
trajectories in between. Note that coherence ratios measure
fluid exchanges between provinces, but do not quantify how
strong the internal mixing is.

The second quantifier we use is a {\sl mixing parameter}
devised to assess how strongly the flow mixes fluid inside
communities. To define the mixing parameter $M_{t_0}^\tau(A)$
inside a set $A$ we first define a transport matrix conditioned
to represent just the transport occurring inside $A$ (more
precisely, transport by trajectories that start and end in
$A$):
\BE
\textbf{R}(t_0,\tau|A)_{ij}=\frac{\mP(t_0,\tau)_{ij}}{\sum_{k
\in \mathcal{I}} \mP(t_0,\tau)_{ik}} \ , \ \ \ i,j \in
\mathcal{I}  \ .
\EE
As before, $\mathcal{I}$ is the set of indices identifying the
boxes $B_i$ making $A$. The mixing parameter is a normalized
version of the sum inside $A$ of the entropies associated to
the transition probabilities in $\textbf{R}(t_0,\tau|A)$:
\BE
M_{t_0}^\tau(A)=\frac{-\sum_{i,j \in \mathcal{I}}
\textbf{R}(t_0,\tau|A)_{ij} \log \textbf{R}(t_0,\tau|A)_{ij}}
{Q_A \log Q_A} \ .
\EE
$Q_A$ is the number of boxes in $A$. The maximum value,
$M_{t_0}^\tau(A)=1$, Is reached when fluid is dispersed from
each box in $A$ to all the others uniformly
($\textbf{R}_{ij}=1/Q_A, \forall i,j\in \mathcal{I}$). A global
quantification of the internal mixing in a community partition
${\cal P}=\{A_1,...,A_p\}$ is given by
\BE
M_{t_0}^\tau({\cal P})=\frac{\sum_{k=1}^p m(A_k)
M_{t_0}^\tau(A_k)}{\sum_{k=1}^p m(A_k)}
\EE
Here, we have weighted the different communities according to
their size.

\subsubsection{Communities in the Mediterranean surface flow}
\label{MediterraneanCommunities}

\begin{figure}
\includegraphics[width=\columnwidth, clip=true]{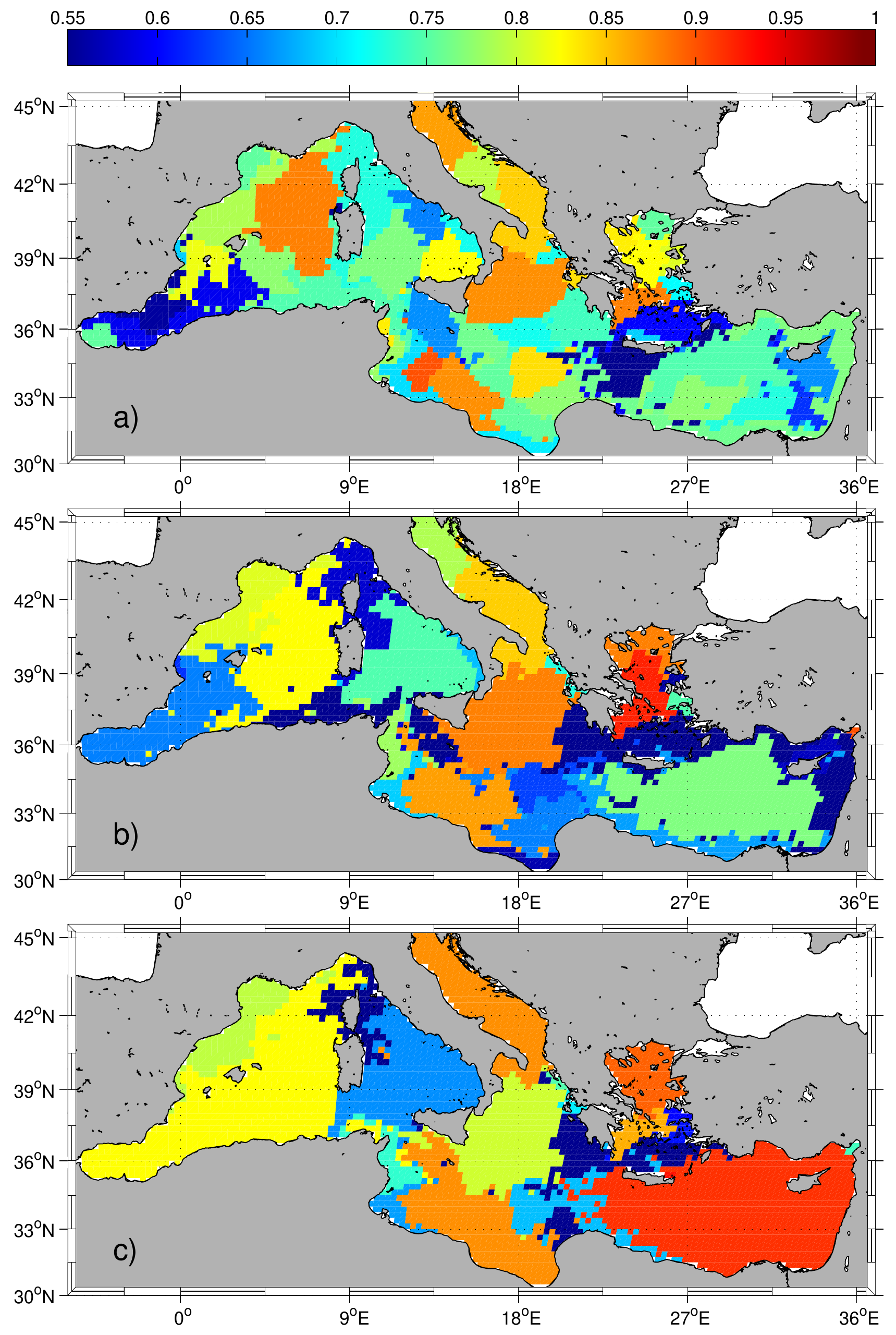}
\caption{\textit{Infomap} partition of flow networks in the Mediterranean sea, defined by $\mP(t_0,\tau)$,
into communities or provinces for increasing values of $\tau$. Each province
is colored by its coherence ratio value from Eq.
(\ref{coherence}), as given in the color bar. In all panels $t_0=$ July 1st 2011.
a) $\tau=30$ days; the number of communities is $p=56$, the global coherence
$\rho_{t_0}^\tau({\cal P})=0.76$, and the global mixing $M_{t_0}^\tau({\cal P})=0.47$. b) $\tau=60$ days; $p=33$,
$\rho_{t_0}^\tau({\cal P})=0.73$, $M_{t_0}^\tau({\cal P})=0.54$. c) $\tau=90$ days; $p=22$,
$\rho_{t_0}^\tau({\cal P})=0.80$, $M_{t_0}^\tau({\cal P})=0.59$.} \label{fig:CommsVsTau}
\end{figure}

The outputs of the \textit{Infomap} algorithm applied to the
flow network defined by $\mP(t_0,\tau)$, for increasing values
of $\tau$, are shown in Fig. \ref{fig:CommsVsTau}. Each
community $A_k$ is colored with the value of its coherence
ratio $\rho_{t_0}^\tau(A_k)$. We see that most coherence values
are rather high. The global mixing parameter has only moderate
values (see caption of Fig. \ref{fig:CommsVsTau}), but it
increases with $\tau$. The main coast-constrained regions
appear clearly outlined (the Tyrrhenian, the Adriatic, the
Aegean, ...), but also other areas defined only by persistent
circulation patterns (the three-gyre system in the Adriatic,
the Balearic front, ...). We refer to \citet{Rossi2014} for a
thorough interpretation of the hydrodynamic provinces in
relation with surface circulation patterns and known
eco-regionalization of the Mediterranean basin. Note that there
is no obvious relationship between the size of a community and
its coherence. Both large and small provinces may have indeed
moderate ($<0.6$) or high ($>0.8$) coherence ratios. The
detection of small communities confirms that \textit{Infomap}
is not affected by the ``resolution limit"\cite{Fortunato2007}.

\begin{figure}
\includegraphics[width=\columnwidth]{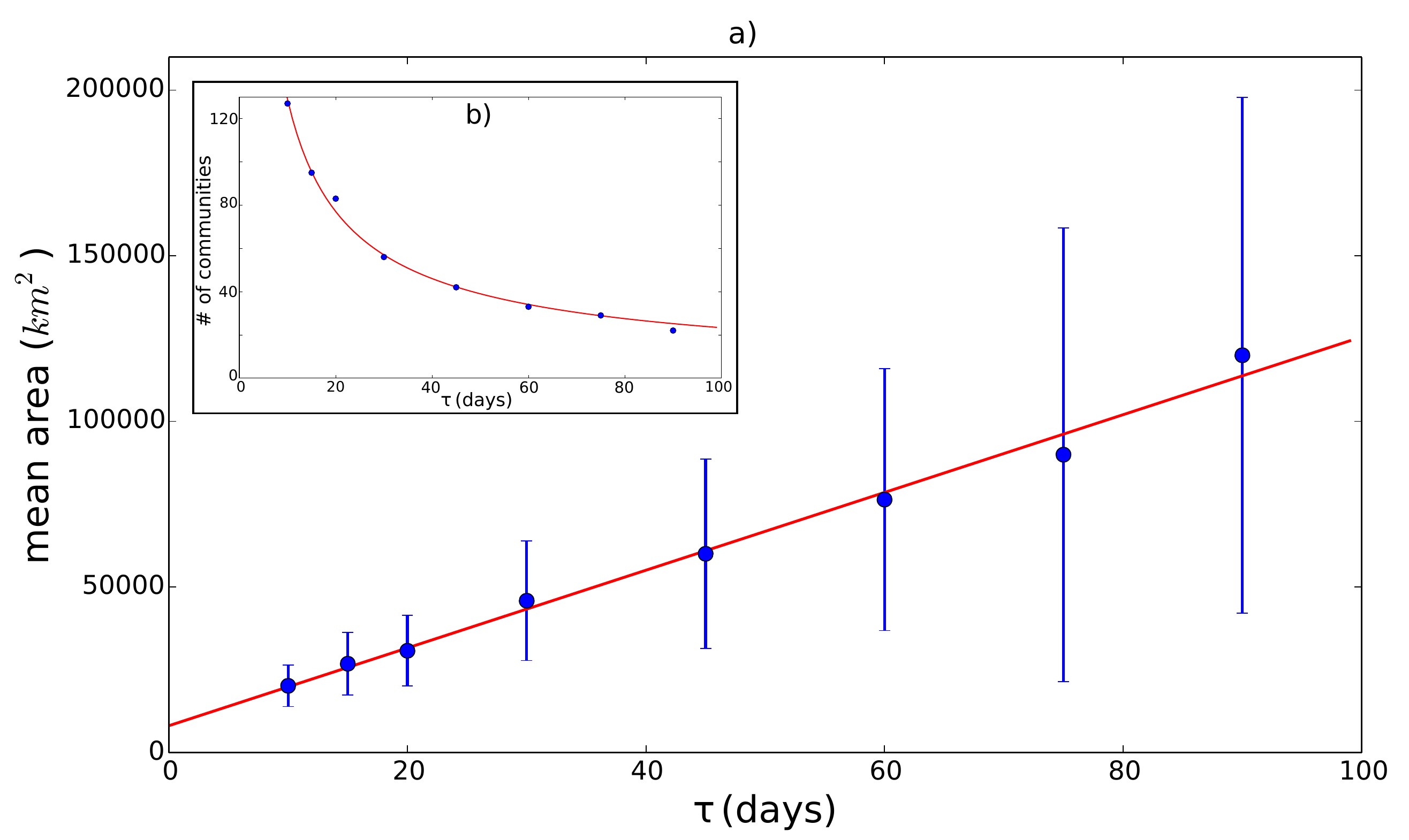}
\caption{Panel a) shows the mean area (dots) of the communities detected by the \textit{Infomap}
algorithm for $t_0=$ July 1st 2011 as a function of $\tau$. The straight line is a
fit to the diffusive-growth-like relationship $\text{Area} = 8109.6 + 1173.8 * \tau$. The error
bars indicate the standard deviation of the area distribution. Note the large dispersion
in community sizes. The upper left inset b) shows the decay of the number of communities with
$\tau$.}
\label{fig:Ncomms}
\end{figure}

Communities merge and in average become larger with increasing
$\tau$, so that their number decreases. Fig. \ref{fig:Ncomms}a
shows the growth of the mean area as a function of $\tau$ for
the same case $t_0=$ July 1st 2011 shown in Fig.
\ref{fig:CommsVsTau}. The standard deviation of the area
distribution is also displayed as error bars. It shows a
significant dispersion in the area of the communities
identified, especially for larger $\tau$, revealing properly
the multi-scale character of oceanic transport processes. For
small $\tau$, community areas seem mainly controlled by the
time of integration (there is no sufficient time for the flow
to manifest highly inhomogeneous dispersion) but only
marginally determined by the intrinsic properties of the flow.
As commented above, detecting communities of widely different
sizes is a great capability of Infomap, whereas other
methodologies constrain the communities to be of similar sizes.
The inset Fig. \ref{fig:Ncomms}b shows how the number of
communities decreases when $\tau$ increases.

\subsubsection{Average descriptions}
\label{Climatological}

\begin{figure}
\includegraphics[width=\columnwidth,clip=true]{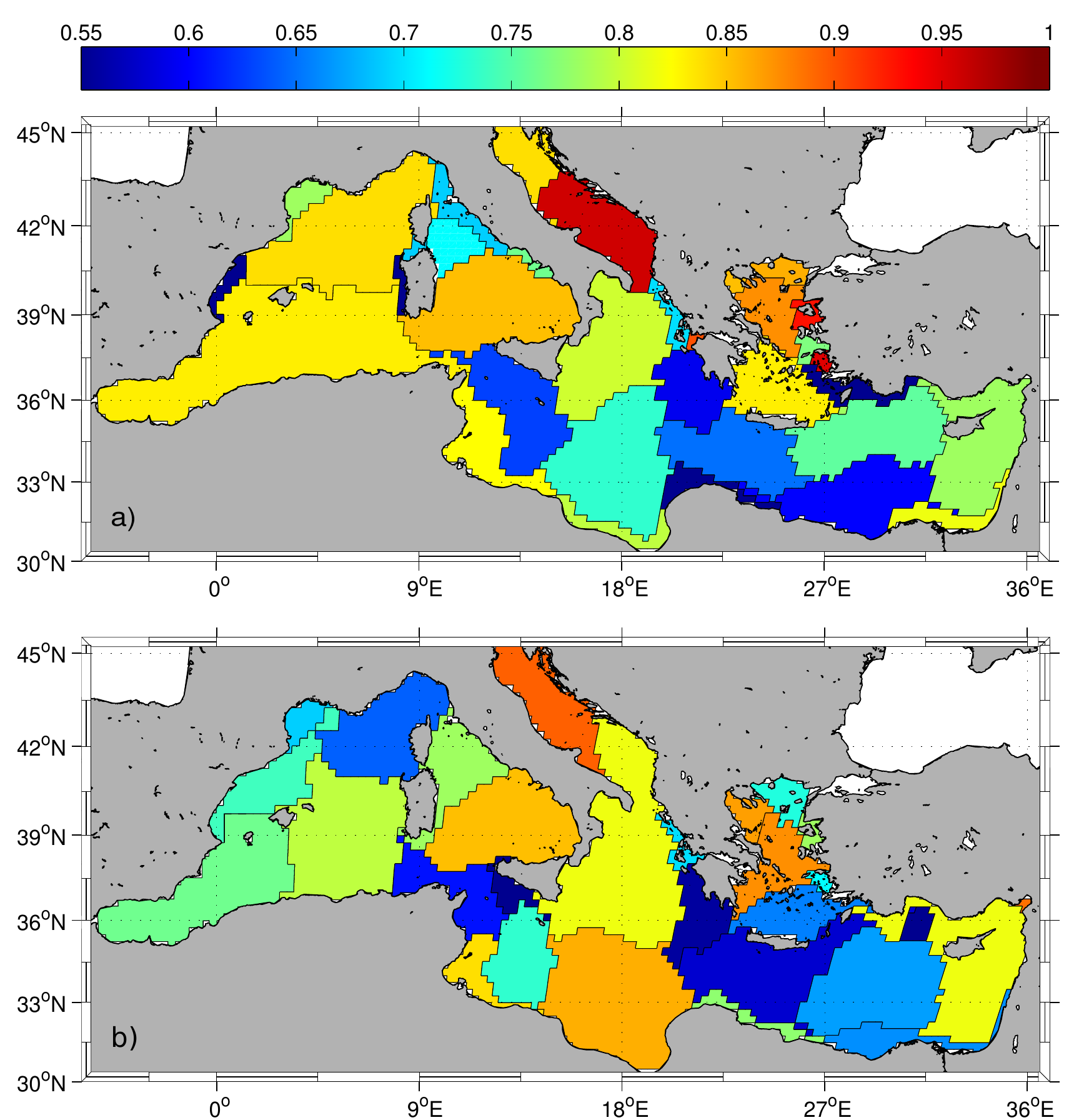}
\caption{\textit{Infomap} communities obtained from the average networks given by
$\overline{\mP(t_0,\tau)}$, with $\tau=30$ days. Each community is colored by
its coherence ratio. a) The average is over the
10 matrices corresponding to $t_0=$ January 1st in 10 years (2002-2011) of simulation;
the number of communities is $p=34$, the global coherence
$\rho_{t_0}^\tau({\cal P})=0.78$, and the global mixing $M_{t_0}^\tau({\cal P})=0.68$.  b)
The average is over the 10 matrices corresponding to $t_0=$ July 1st in the 10 years 2002-2011;
$p=30$, $\rho_{t_0}^\tau({\cal P})=0.77$, $M_{t_0}^\tau({\cal P})=0.69$.
}
\label{fig:average_matrix}
\end{figure}

Because of the turbulent nature of oceanic motions, the
community decomposition changes with $t_0$. Some communities
(even of small size) are repeatedly observed while some others
appear and disappear when changing $t_0$. In order to identify
persistent communities, i.e. those whose limits are relatively
stable in space and time, we explore two averaging procedures
leading to a mean -``climatological"- community partition. In a
first approach we average a number of matrices $\mP(t_0,\tau)$
corresponding to the same starting date (e.g. January 1st) for
the ten different years of the data set (e.g. January 1st 2002,
January 1st 2003, etc. until January 1st 2011). Figure
\ref{fig:average_matrix} shows the \textit{Infomap} partition
of the network defined by the average matrix
$\overline{\mP(t_0,\tau)}$ made with the ten matrices
$\mP(t_0,\tau)$ using the same starting date for each of the 10
years (2002-2011). An example of $t_0$ in winter and another
one in summer are displayed. The figure shows the most
persistent communities for a particular month, averaging out
the variability occurring over ten years. We remark than some
communities have a rather small size (most of them reflecting
shallow oceanic regions such as continental shelves), and that
there is some inter-seasonal variability.

\begin{figure}
\includegraphics[width=\columnwidth]{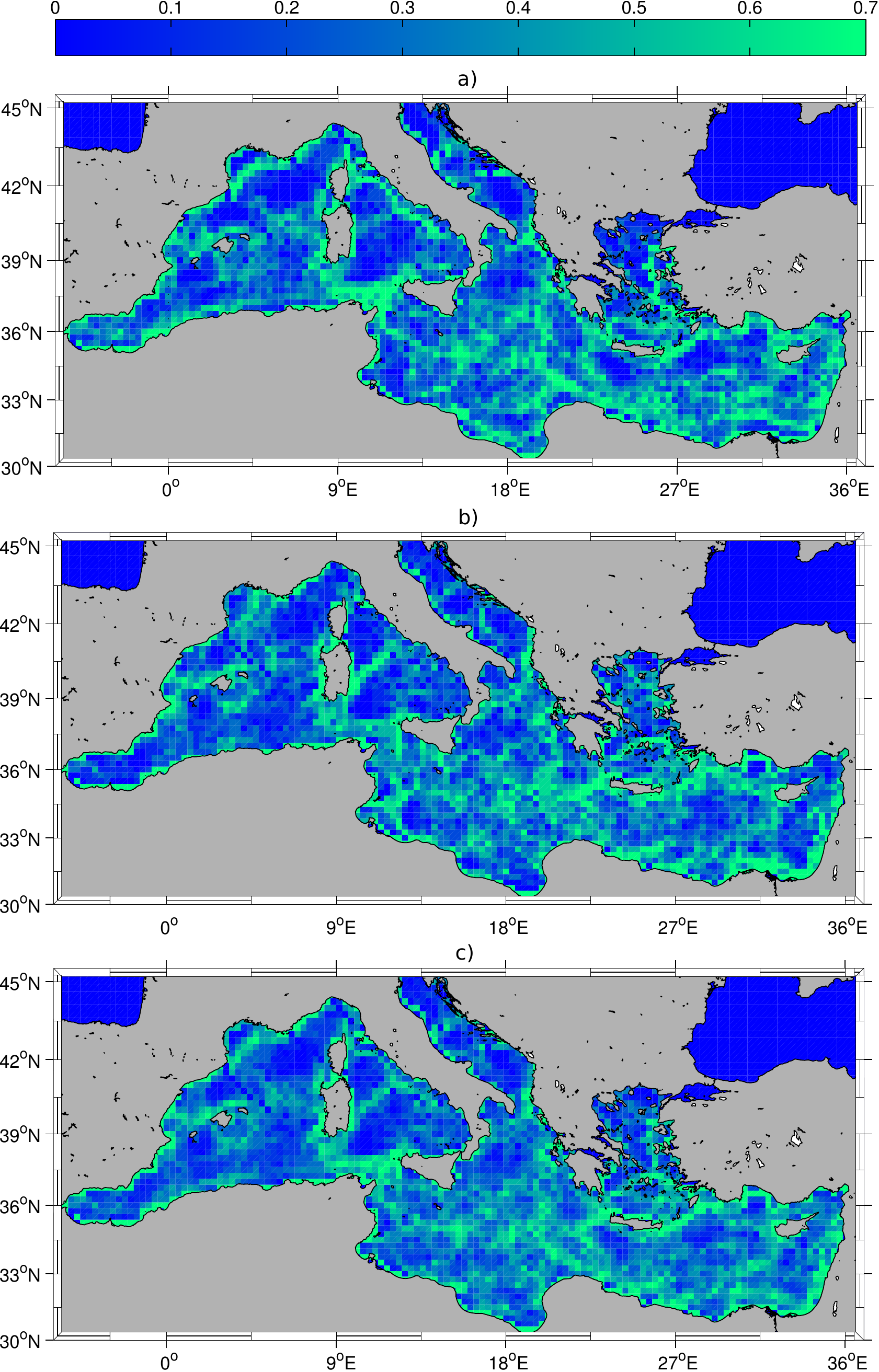}
\caption{Persistence of community borders over time: Color code
indicates the proportion of times one of the borders between communities
has appeared at a given location. $\tau=30$ days.
a) $t_0=$ January 1st of (2002-2011). b) $t_0=$ July 1st of (2002-2011).
c) The average of the
two previous panels, eliminating the seasonal information. }
\label{fig:average_borders}
\end{figure}

A second approach to obtain average or climatological
description of the community partition is illustrated in Fig.
\ref{fig:average_borders}. Instead of applying only once
\textit{Infomap} on an averaged transport matrix, it is here
applied 10 times separately on the 10 transport matrices
corresponding to the same starting date for each of the 10
years (2002-2011). The color at a particular location of Fig.
\ref{fig:average_borders} indicates the frequency of occurrence
(in these 10 partitions) at that location of a border between
communities. Then, greener color indicates a more persistent
community border. The strongest lines would represent true
``barriers to transport" which remain fixed in space. Fuzzier
lines may indicate intermittent border appearance, but also a
larger wandering amplitude. Figures \ref{fig:average_borders}a
and b display the situation in the same winter and summer days
as in Fig. \ref{fig:average_matrix}. Figure
\ref{fig:average_borders}c shows a combination of them,
equivalent to showing the barrier persistence sampled twice a
year during the ten years.

\subsubsection{Comparison with spectral partitioning}
\label{Spectral}

\begin{figure}
\includegraphics[width=\columnwidth,clip=true]{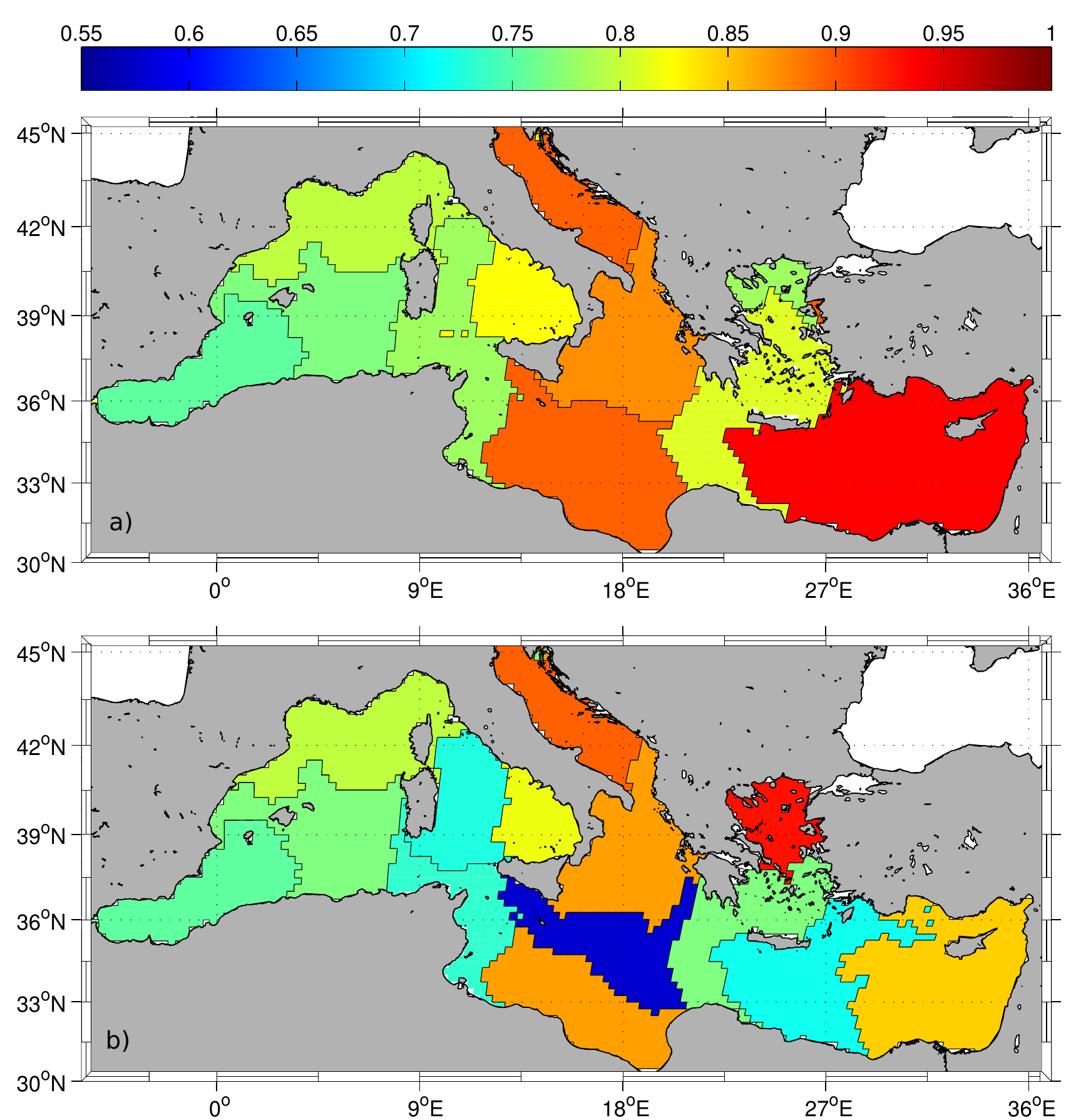}
\caption{Community decomposition by the spectral method with fuzzy c-means
clustering described in \citet{Froyland2003}. The matrix used is the
same average $\overline{\mP(t_0,\tau)}$ as in Fig.
\ref{fig:average_matrix}b), i.e. with $t_0=$ July 1st, averaged in the ten years 2002-2011,
and $\tau=30$ days. Ten eigenvalues are used. a) The number of communities is fixed to
be $p=10$; the global coherence is
$\rho_{t_0}^\tau({\cal P})=0.85$, and the global mixing is $M_{t_0}^\tau({\cal P})=0.62$.
In the Aegean, the southern yellow community is the only independent one: the
portions of the Aegean further north are clustered by
the c-means algorithm as being part of the same province as areas in the central Mediterranean
with the same color. b) $p=14$;
$\rho_{t_0}^\tau({\cal P})=0.78$, $M_{t_0}^\tau({\cal P})=0.64$.  }
\label{fig:spectral}
\end{figure}

Different methods based on the spectral properties of transport
matrices have been previously used to identify and locate
almost-invariant sets in flows
\cite{Froyland2003,Froyland2005,Froyland2007,Dellnitz2009,Speetjens2013}.
They exploit the fact that for a set to remain almost invariant
after the effect of the flow, it has to be related with
eigenvectors of $\mP(t_0,\tau)$  with eigenvalues close to 1.
Here we compare our partitioning obtained by \textit{Infomap}
with the one from those spectral methods. To be specific we
consider the method described by \citet{Froyland2003}. The
technique in this last paper obtains a partition ${\cal P}$
minimizing in an approximate way the global coherence
$\rho_{t_0}^\tau({\cal P})$. To this end it computes
eigenvectors associated to nearly vanishing eigenvalues of the
Laplacian matrix\cite{Newman2010} obtained from the symmetric
part of $\mP(t_0,\tau)$, and combines them using a fuzzy
c-means clustering algorithm \cite{Froyland2003}. Note that
this approach eliminates any directionality information present
in the transport network. Also, the c-means clustering can
define as a single community pieces of the ocean which are
geographically disjoint or in fact quite far apart, if this
enhances the coherence defined in Eq. (\ref{globalcoherence}).
In the method, one has to specify the number of eigenvectors
being combined (we choose it to be 10) and the number of
communities in the partition. Figure \ref{fig:spectral} shows
the results using the same average matrix
$\overline{\mP(t_0,\tau)}$ as in Fig.
\ref{fig:average_matrix}b, and imposing a partition in 10 and
in 14 communities. The change in the number of communities
leads to rearrangements in the Tyrrhenian, the central
Mediterranean, the Aegean, and the Levantine basin. In panel a)
some of the communities are made of disjoint pieces. Larger
number of communities decreases the global coherence ratio (see
caption of Fig. \ref{fig:spectral}). If we try to increase the
number of communities approaching the one given by
\textit{Infomap} we find that the clustering algorithm becomes
unstable. Instabilities also occur when the number of links in
the transport network becomes too high (as occurring for
example when increasing $\tau$ beyond 1 month).

When compared with the \textit{Infomap} decomposition we see
that several of the boundaries coincide. But there are
important differences, such as the wider range of community
sizes and the sharper details revealed by \textit{Infomap}.
This is because a constraint of similar sizes for the
communities associated to the same eigenvector needs to be
imposed in the spectral method. When clustering several
eigenvectors together this limitation is partially bypassed but
still not removed. The values of the coherence ratio are of the
same order or somehow larger for the spectral method, but note
that the number of spectral communities has been kept much
smaller to avoid the instabilities in the clustering algorithm.
Since merging two communities into a single one increases the
global coherence, joining some of the \textit{Infomap}
communities in Fig. \ref{fig:average_matrix} until arriving to
10 or 14 communities as in Fig. \ref{fig:spectral} would give
rather large values of $\rho_{t_0}^\tau({\cal P})$. As
expected, the global mixing parameter is larger for the
\textit{Infomap} partition, but only by a small amount,
reflecting that, even if internal mixing is not imposed in the
spectral method, it is achieved to a reasonable extent.

From the methodological point of view, \textit{Infomap}
presents the advantage of determining itself the number of
communities in the partition, whereas this needs to be fixed
\textit{a priori} (as well as the number of eigenvectors to be
clustered) in the spectral approach. On the other hand, the
spectral method is formulated as an algorithm to minimize the
global coherence ratio, a quantity with a clear physical
meaning. The quantity optimized by \textit{Infomap} is a
codeword length given in Eq. (\ref{map}) of Appendix B, an
abstract information-theoretic object without a clear physical
meaning. The heuristic interpretation of the optimization
process leads to the `large internal-small external
connectivity' property for the communities, but a more rigorous
understanding of the \textit{Infomap} procedure is clearly
needed\cite{Rosvall2008,Schaub2012}.

The results of this section indicate that the \textit{Infomap}
methodology proposed here to identify coherent fluid regions
seems more appropriate than spectral methods when a wide range
of community sizes is expected, when internal mixing is a key
parameter, or to minimize user input (such as entering the
number of communities). Spectral methods seem appropriate when
one is looking precisely for the sets defined mathematically as
almost-invariant, the coherence ratio describes well the
desired properties of the partition, and one expects a limited
range of sizes.


\section{Conclusions}
\label{sec:conclusions}

We have used the concept of {\sl flow networks} to obtain a
discretized view of transport processes in geophysical
contexts. Once the fluid motion is cast into the graph-theory
framework, powerful techniques from this field become available
to investigate the fluid transport processes. In this paper we
have improved, using network concepts, the characterization of
geophysical dispersion and mixing process, as well as the
identification of coherent fluid regions. One of the simplest
network descriptors, the degree of a node, gives direct
information on local stretching properties, classically
associated to the finite-time Lyapunov exponents and their
distributions. Thus the out-degree at a particular node is
quantitatively related to the fluid stretching at that location
in the time-forward direction, and quantifies fluid dispersion.
The in-degree is related to backwards stretching, and thus to
the mixing of fluid from different origins.

A family of {\sl network-entropy} functions has been defined,
aiming at describing higher order statistical properties of
fluid stretching (and then of dispersion and mixing) in terms
of the network adjacency matrix. One of them, $H_i^0$ is simply
the logarithm of the degree. Another one, $H_i^1$, is the
discrete finite-time entropy studied by
\citet{Froyland2012Finite}. We find numerically that it
provides a good estimation of the coarse-grained finite-time
Lyapunov exponent. We expect higher order entropies to be
related to the generalized Lyapunov exponents
\cite{Boffetta2002,Cencini2010} that characterize successive
moments of the Lyapunov field. Further work is needed to assess
the validity of these properties more rigourously, beyond the
heuristic and numeric arguments given in Sect.
\ref{subsec:network:dispersion} and in Appendix A.

This paper considered flow networks in the geophysical context,
but it is anticipated that the concepts are equally valid in
more general fluid dynamics context, and even apply to more
abstract flows in the phase space of dynamical systems
\cite{Froyland2003,Dellnitz2006,Santitissadeekorn2007}. Also,
the network entropies defined here can be used to characterize
the local properties of general weighted networks beyond the
degree and the node strength.

As a second application in which the network representation
provides useful insights we have investigated the
identification of coherent regions in the ocean
flow\cite{Nilsson-Jacobi2012,Thomas2014,Rossi2014}, regions
that are similar to {\sl almost invariant}
sets\cite{Froyland2003,Froyland2005} but for which the presence
of strong internal mixing is also desired. We find in the
network-theory toolbox a useful community detection technique,
\textit{Infomap}, that takes into account the directed and
weighted nature of the flow network, and that finds partitions
of the geophysical flow with the required characteristics
without restricting the range of community sizes. We have
argued that these characteristics make it an interesting
alternative to spectral methods to identify the desired
coherent regions, although we also recognize that a substantial
clarification of the physical meaning of the minimization
process involved in \textit{Infomap} is needed. This partition
of the sea into coherent {\sl provinces} has already been used
to evaluate larval connectivity and to inform the design of
marine protected areas\cite{Rossi2014}. The present
implementation of the method deals only with regions fixed with
respect to geographic boundaries. Tools from the study of
bipartite networks would be needed to find moving coherent
regions such as
vortices\cite{Santitissadeekorn2010,Froyland2012Three}.

We believe that the representation of fluid motion as a
transport or flow network, allowing the use of powerful
techniques from graph theory, will continue to provide novel
insights into the nonlinear processes occurring in our planet,
most of them related to fluid transport.

\begin{acknowledgments}
We acknowledge financial support from FEDER and MINECO (Spain)
through the ESCOLA (CTM2012-39025-C02-01) and INTENSE@COSYP
(FIS2012-30634) projects, and from European Commission
Marie-Curie ITN program (FP7-320 PEOPLE-2011-ITN) through the
LINC project (no. 289447). The simulated velocity field used
here was generated by MyOcean (http://www.myocean.eu/). The
authors thank two anonymous reviewers for their constructive
comments.
\end{acknowledgments}

\section*{Appendix A: Relationship between network entropies and stretching statistics}
\renewcommand{\theequation}{A\arabic{equation}}
\setcounter{equation}{0}  
\renewcommand{\thefigure}{A\arabic{figure}}
\setcounter{figure}{0}  

\begin{figure}
\includegraphics[width=\columnwidth,clip=true]{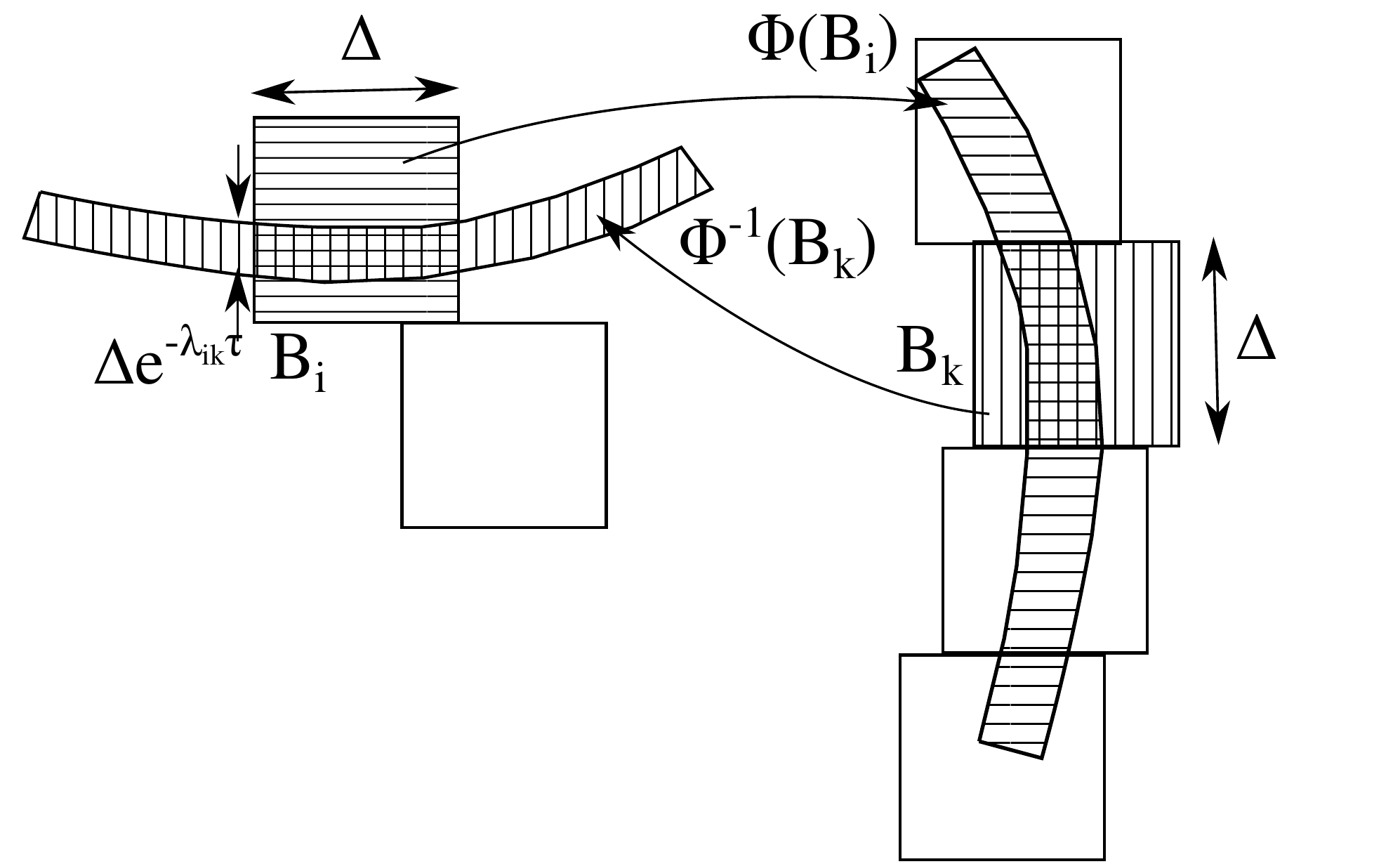}
\caption{Schematics of the stretching (forward and backwards in time) of fluid boxes
of sidelength $\Delta$ corresponding to
network nodes. $\Phi(B_i)$ is a shortcut for $\Phi_{t_0}^\tau(B_i)$, and $\Phi^{-1}(B_k)$
is a shortcut for $\Phi_{t_0+\tau}^{-\tau}(B_k)$. $\lambda_{ik}$ is the value of the forward FTLE
$\lambda(\bx_0,t_0,\tau)$ in the doubly dashed region $B_i \cap \Phi^{-1}(B_k)$. }
\label{fig:appendixA}
\end{figure}

In this Appendix we derive heuristically relationships between
the network entropies defined in Sect.
\ref{subsec:network:dispersion} and Lyapunov exponent
statistics (in the two-dimensional case).  Fig.
\ref{fig:appendixA} illustrates the basic ideas. The
assumptions are that dynamics is mainly hyperbolic in the
region of interest, and that $\tau$ and the size $\Delta$ of
the fluid boxes $\{B_i, i=1,...,N\}$ are such than the image of
the boxes by the flow after a time $\tau$ are thin and long
filaments. Boxes in the partition have been roughly aligned
with expanding and contracting directions to make easier the
heuristic arguments.

The point is to estimate the values of the matrix elements
$\mP(t_0,\tau)_{ik}$ given in Eq. (\ref{PF}):
\BE
\mP(t_0,\tau)_{ik} = \frac{m\left(B_i \cap
\Phi_{t_0+\tau}^{-\tau}(B_k)\right)}{m(B_i)} \ .
\label{PFappendix}
\EE
The quantity in the numerator of Eq. (\ref{PFappendix}) is the
area of the doubly-dashed thin filament in the left of Fig.
\ref{fig:appendixA}. If we assume that the forward FTLE
$\lambda(\bx_0,t_0,\tau)$ is approximately constant for $\bx_0$
in this region, we have $m\left(B_i \cap
\Phi_{t_0+\tau}^{-\tau}(B_k)\right) \approx
\Delta^2\exp(-\lambda_{ik}\tau)$ (see Fig.
\ref{fig:appendixA}), where $\lambda_{ik}$ is this constant
value. In consequence,
$\mP(t_0,\tau)_{ik}\approx\exp(-\lambda_{ik}\tau)$ if $B_k$ is
one of the boxes containing part of the image
$\Phi_{t_0}^\tau(B_i)$ of $B_i$, and $\mP(t_0,\tau)_{ik}=0$
elsewhere.

Spatial features in typical forward FTLE fields are thin
filaments with nearly constant value $\lambda$. They are
elongated along the expanding
directions\cite{Haller2001,Shadden2005} and have widths of the
order of $l\exp(-\lambda\tau)$, where $l$ is the size of the
velocity field inhomogeneities, i.e. the size of the Eulerian
structures driving the flow. Then, the uniformity condition we
are imposing is $\Delta<l$, i.e. discretization boxes smaller
than Eulerian structures. In our Mediterranean example,
$\Delta$ is smaller than the dominant mesoscale structures in
the sea, but some of the smaller features in the velocity field
can have some impact on the validity of the uniformity
condition.

We can use our estimation of $\mP(t_0,\tau)_{ik}$ to compute
the sum appearing in the network entropies definition Eq.
(\ref{Hiq}). The assumption of uniform FTLE inside region $B_i
\cap \Phi_{t_0+\tau}^{-\tau}(B_k)$ allows us to freely replace
functions of $\lambda_{ik}$ by average values in that region:
\BA
&&\sum_{k=1}^N \left(\mP(t_0,\tau)_{ik}\right)^q \approx
\sum_{k=1}^N e^{-q\lambda_{ik}\tau}\approx \nonumber \\
&&\sum_{k=1}^N \frac{1}{\Delta^2
e^{-\lambda_{ik}\tau}}\int_{B_i \cap
\Phi_{t_0+\tau}^{-\tau}(B_k)}
e^{-q\lambda(\bx_0,t_0,\tau)\tau}d\bx_0 \approx \nonumber \\
&& \sum_{k=1}^N \frac{1}{\Delta^2}\int_{B_i \cap
\Phi_{t_0+\tau}^{-\tau}(B_k)}
e^{(1-q)\lambda(\bx_0,t_0,\tau)\tau}d\bx_0 = \nonumber \\
&& \hspace{2.5cm} \frac{1}{\Delta^2}\int_{B_i}
e^{(1-q)\lambda(\bx_0,t_0,\tau)\tau}d\bx_0 \ ,
\EA
which, using definition (\ref{Hiq}), implies
\BE
e^{(1-q)\tau H_i^q(t_0,\tau)} \approx \left<
e^{(1-q)\tau\lambda(\bx_0,t_0,\tau)}\right>_{B_i} \ .
\label{expHiq}
\EE
This is the sought relationship between network entropies and
moments of the stretching factor $e^{\lambda\tau}$. For $q=0$
we reobtain Eq. (\ref{expHi0}). In the limit $q\rightarrow 1$
we get $H_i^1(t_0,\tau) \approx \left<
\lambda(\bx_0,t_0,\tau)\right>_{B_i}=\lambda_i(t_0,\tau)$. The
arguments above can be repeated to get the same relationship
(\ref{expHiq}) between network entropies in the backwards time
direction and backwards Lyapunov exponents.

All these expressions are similar to the ones presented for
example by \citet{Paladin1987} relating R\'{e}nyi entropies and
generalized Lyapunov exponents defined from moments of the
stretching factor $e^{\lambda\tau}$. But here the moments are
not by averaging along a dynamic trajectory but inside a box
$B_i$. In the same way as the value of any of the network
entropies at node $i$ characterizes the inhomogeneity in the
fluxes sent from $i$ to other nodes, the difference between the
different entropies (different $q$) at a single node $i$
characterizes the inhomogeneity of the FTLE inside box $B_i$.
This is a way by which small-scale features present in the
Lagrangian trajectories get statistically represented in the
network description. Relationships such as (\ref{expHiq}) are
not exact for finite $\Delta$ and $\tau$, but we expect them to
become more accurate for increasing $\tau$ and decreasing
$\Delta$.

\section*{Appendix B: The \textit{Infomap} method}
\renewcommand{\theequation}{B\arabic{equation}}
\setcounter{equation}{0}  

\textit{Infomap} \cite{Rosvall2008} is a community-detection
algorithm
\cite{Newman2010,Danon2005,Lancichinetti2009,Fortunato2010,Aldecoa2013}
that retains both the ``direction'' and ``weight'' information
of each link in the network.

\textit{Infomap} does not require to specify \emph{a priory}
the number of communities to be detected. It finds structures
which are directly related to well-mixed regions under the flow
represented by $\mP(t_0,\tau)$, and not to other structural
properties (for example, a well defined region with strong
fluxes oriented towards a particular direction) which will not
lead to particle localization in that region. Also,
\textit{Infomap} does not assume communities with similar sizes
(as for example spectral
partitioning\cite{Froyland2005,Froyland2007}) nor suffers from
the `resolution limit' \citep{Fortunato2007} which limits the
minimum community size detectable by most algorithms. In fact,
the method decomposes the transport network into subgraphs of
different sizes where the flow requires so.

In addition to these convenient properties, the minimization
algorithm is efficiently implemented in publicly available
software (http://www.tp.umu.se/\verb+~+rosvall/code.html).

\textit{Infomap} considers an ensemble of random walkers in the
weighted and directed network defined by $\mP(t_0,\tau)$,
moving with the transition probabilities in that matrix. Then,
the method considers from the information-theory point of view
the optimal coding of the ensemble of possible random walks. To
this end the network is divided in communities and each random
walk is coded by sequences of words that represent successive
locations inside a community and jumps between different
communities. The information-theoretic lower bound to the
average length of the codeword used is given in terms of the
transition probabilities and of the specific partition in
communities by the so-called {\sl map equation}:
\begin{equation}
L=q_\curvearrowright  H({\cal Q}) + \sum_{\alpha=1}^c p_\circlearrowright^\alpha H({\cal P^\alpha})\ .
\label{map}
\end{equation}
$c$ is the number of communities in the particular partition
considered. The first term involves the Shanon entropy
associated to the transitions between different communities
$\alpha$:
\begin{equation}
H({\cal Q})= -\sum_{\alpha=1}^c \frac{q_{\alpha\curvearrowright}}{q_\curvearrowright}
\log_2\left(\frac{q_{\alpha\curvearrowright}}{q_\curvearrowright}\right)
\end{equation}
$q_{\alpha\curvearrowright}$ is the probability to leave
community $\alpha$ in one random-walk step, and
$q_\curvearrowright=\sum_{\alpha=1}^c
q_{\alpha\curvearrowright}$. Expressions for these quantities
in terms of the components of the network matrix
$\mP(t_0,\tau)$ are given in \citet{Rosvall2008}. The second
term in Eq. (\ref{map}) contains the Shanon entropies $H({\cal
P^\alpha})$ associated to the words used to codify the position
inside a community $\alpha$ and the word that denote the exit
from that community:
\begin{equation}
H({\cal P^\alpha})= -\sum_{i\in\alpha} \frac{\pi_i}{p_\circlearrowright^\alpha}
\log_2\left(\frac{\pi_i}{p_\circlearrowright^\alpha}\right)- \frac{q_{\alpha\curvearrowright}}{p_\circlearrowright^\alpha}
\log_2\left(\frac{q_{\alpha\curvearrowright}}{p_\circlearrowright^\alpha}\right)\ .
\end{equation}
The notation $i\in\alpha$ indicates sum over the nodes
pertaining to community $\alpha$. $\pi_i$ is the stationary
distribution of the random walk and
$p_\circlearrowright^\alpha=q_{\alpha\curvearrowright}+\sum_{i\in\alpha}
\pi_i$. Again, expressions for these quantities can be obtained
from the elements in the network matrix $\mP(t_0,\tau)$
\citep{Rosvall2008}.

\textit{Infomap} finds the partition that minimizes the
quantity in (\ref{map}), i.e. the partition that provides a
shorter description of the ensemble of walks going in and
outside the communities. In other words, it finds the partition
for which the random walks remain most of the time inside the
communities with few jumps between them. This minimization
process uses a deterministic greedy algorithm followed by a
simulated-annealing which was repeated $100$ times to select
the best partition in provinces (although the results were
already stable after $10$ attempts).


%

\end{document}